\documentclass[aps,amsfonts,pre,twocolumn,superscriptaddress,showpacs,eqsecnum]{revtex4-1}
\usepackage{epsfig,amsmath,amssymb,bm,epsf,graphics,psfrag,verbatim,subfigure}

\usepackage[normalem]{ulem}
\usepackage{float}
\usepackage[usenames]{color}
\usepackage{graphicx}
\usepackage{ulem}
\usepackage{amsfonts}
\usepackage{amsmath}
\usepackage{natbib}
\usepackage{epsfig,amsmath,amssymb,bm,epsf,graphics,psfrag,verbatim,subfigure}

\def\hoa{\phi}
\def\vq{\vec{q}}
\def\uvepara{\hat{e}^{\parallel}}
\def\uveperp{\hat{e}^{\perp}}
\def\g{\gamma}

\def\vr{\mathbf{r}}
\def\ori{\hat{n}}
\def\uve{\hat{e}}
\def\uvm{\hat{m}}

\def\HCF{H_{\textrm{CF}}}
\def\HEF{H_{\textrm{Effective}}}
\def\MS{\Omega}
\def\hps{\phi}

\begin{document}

\title{Entropic effects in the self-assembly of open lattices from patchy particles}

\author{Xiaoming Mao}
\affiliation{Department of Physics, University of
Michigan, Ann Arbor, MI 48109, USA }

\date{\today}
\begin{abstract}
Open lattices are characterized by low volume-fraction arrangements of building blocks, low coordination number, and open spaces between building blocks.  The self-assembly of these lattices faces the challenge of mechanical instability due to their open structures.  We theoretically investigate the stabilizing effects of entropy in the self-assembly of open lattices from patchy particles.  A preliminary account of these findings and their comparison to experiment was presented recently [X. Mao, Q. Chen, S. Granick, \textit{Nat. Mater.}, \textbf{12}, 217 (2013)].  We found that rotational entropy of patchy particles can provide mechanical stability to open lattices, whereas vibrational entropy of patchy particles can lower the free energy of open lattices and thus enables the selection of open lattices verses close-packed lattices which have the same potential energy.  These effects open the door to significant simplifications of possible future designs of patchy-particles for open-lattice self-assembly.

\end{abstract}


\maketitle

\section{Introduction}
Open structures, structures characterized by low volume-fraction arrangements of building blocks with open spaces between them, are ubiquitous in both natural~\cite{Fletcher2010,Weiner1998,Davis1992,Yurchenco1990,Hammonds1996,Witten1981,Schaefer1986} and engineered~\cite{Horike2009,Davis2002,Lee2009,Schaedler2011,Lee2001} systems.  Open-structure materials occur in a rich variety of morphologies, and exhibit fascinating aspects in their mechanical~\cite{Sun2012,Kapko2010,Sartbaeva2006,Greaves2011}, transport~\cite{Davis1992,McDowellboyer1986}, and thermodynamical properties~\cite{Hammonds1996,Ernst1998}, leading to important applications in engineering.  For example, zeolite~\cite{Davis1992}, a natural aluminosilicate mineral, consisting open-lattice structure with pores of sizes at the nanometer scale, has been widely used as catalyst, molecular sieves, solar thermal collectors, etc., and has been shown to exhibit a number of unusual features including negative thermal expansion~\cite{Hammonds1996}, negative Poisson’s ratio~\cite{Greaves2011,Grima2007,Grima2000}, and flexibility window~\cite{Sartbaeva2006}.

Structural openness of a material often leads to structural flexibility, which is a double-edged sword.  On the one hand, open structures are often close to mechanical instability point and this poses great challenges to the fabrication of open-structure materials.  On the other hand, the closeness to mechanical instability point also opens the door to interesting designs of novel materials with interesting applications such as precisely controlled, fast, and reversible deformations, etc.

The mechanical instability of open-structure materials can be understood through a simple counting argument due to J.~C.~Maxwell in 1864~\cite{Maxwell1864}.  The basic idea of this counting argument is that to determine whether a system is mechanically stable one can simply compare the total number of degrees of freedom $N_d$ and total number of constraints $N_c$.  If $N_d>N_c+d(d+1)/2$, where $d(d+1)/2$ represent the degrees of freedom of the global translations and rotations of the whole system, this system does not have sufficient number of constraints to fix all the relative degrees of freedom and there exist deformations, i.e., ways to rearrange the different components of the system, without violating any of the constraints.  These deformations are called \lq\lq floppy modes\rq\rq and we use $N_0$ to denote the number of these floppy modes.  The point at which $N_d$ and $N_c+d(d+1)/2$ (assuming all constraints non-redundant) becomes equal defines a mechanical critical point, at which the system is at the onset of mechanical stability.  This point is usually called the \lq\lq isostatic\rq\rq  point.  Above this point, the system exhibit mechanical stability and there are no floppy modes.  
For an actual system, constraints often take the form of elastic connections, e.g., contacts between colloidal particles, sections of semi-flexible polymers between crosslinks, struts in frames, or chemical bonds between atoms.  Correspondingly, floppy modes take the form of zero-energy deformations, which lead to mechanical instabilities in some form.  Other non-floppy modes in the system are allowed but will cost elastic energy.  

In particular, for periodic lattice structures consisting of point-particles with only nearest neighbor central-force interactions, this criterion of no floppy modes can be simplified by ignoring the global translational/rotational degrees of freedom $d(d+1)/2$ given that it is much smaller than $N_d$ and $N_c$ in a large lattice.  Thus one have the simple equation for isostaticity in periodic lattices, $z=2d$, where $z$ is the coordination number, i.e., the number of connecting neighbors of each particle.  At this point, each lattice site has just enough constraints to fix all its degrees of freedom.  In practice, lattices are of finite size, and particles on the boundary have fewer constraints than the ones in the bulk.  As a result there exist a subextensive number of floppy modes in a finite lattice in which $z=2d$ in the bulk.  These floppy modes can either take the form of extended or surface floppy modes, depending on the geometry of the lattice~\cite{Sun2012}.

Periodic open-structure lattices, such as honeycomb, kagome lattices in two dimensions and diamond, pyrochlore, perovskite, diamond lattices in three dimensions, are typically either at or below the isostatic point, thus at finite size they exhibit floppy modes which lead to mechanical instability and makes them challenging to produce in practice, as we mentioned above.  In existing open-structure materials, additional interactions beyond nearest-neighbor central-force interactions provide more constraints which stabilize the open lattices.  For example, in open-lattice crystals, covalent bonds with preference on bond-angle directions stabilizes spatially open structures such as the zeolite or crystobalite, whereas in bio-polymer networks the molecular structure of the polymers provides stiffness against bending and thus provides mechanical stability.

At the colloidal scale, these stabilization mechanisms are difficult to realize, because interactions are typically nearest-neighbor without bond-angle selection.  Provided the wide predicted applications of open structures at this scale in fields such as catalysis and photonics~\cite{Joannopoulos1997,Moroz2002,Garcia-Adeva2006,Galisteo-Lopez2011,Chen2011}, many designs have been proposed to obtain open-lattices from self-assembly in the colloidal scale via additional \emph{potential energy} terms such as further range interaction, three- or more-body interaction, DNA-functionalization, etc.~\cite{Cohn2009,Edlund2011,Torquato2009,Velikov2002,Glotzer2007,Romano2012,Tkachenko2002}.  Most of these designs are still difficult to realize given current experimental techniques on colloidal particles.  

In this Paper, we discuss a new mechanism of stabilizing open colloidal lattices through entropic effects in patchy particles, and present in detail a systematic theory that captures this effect and provides guidelines to remarkably simple future designs of open-lattice self-assembly.  We apply this theory to the model system of a two-dimensional self-assembly of triblock Janus particles which are particles with attractive patches on the north- and south- polar caps~\cite{Chen2011,Mao2013}, and arrive at free energies of different lattice geometries and the equilibrium phase diagram under pressure.  This theory is readily generalizable to other systems of self-assembly of patchy particles.

\begin{figure}
	\centering
		\includegraphics[width=.25\textwidth]{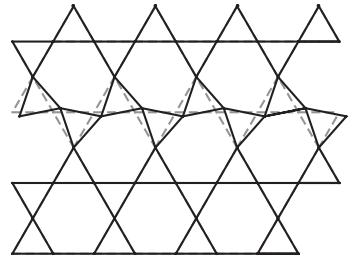}
	\caption{The kagome lattice and one example of its floppy modes.  The floppy-mode deformed kagome lattice is shown in solid lines, and the undeformed lattice is in dashed lines.  The deformations only occur in a horizontal line in which the pointing-up and pointing-down triangles are rotated.  In other parts of the lattice the deformed and the undeformed lattices are on top of each other.
	}
	\label{FIG:FM}
\end{figure}

Using the Maxwell's counting argument discussed above, the two-dimensional kagome lattice is at the isostatic point.  At finite size with free boundary condition, there exist a sub-extensive number of floppy modes.  For the kagome lattice these floppy modes correspond to the rotations of triangles along straight lines in the lattice as shown in Fig.~\ref{FIG:FM}.  These modes are of zero potential-energy cost if there are only nearest-neighbor central-force interactions in the lattice, because to leading order they keep the length of all bonds unchanged.  Finite amplitude floppy modes also exist in the kagome lattice and are discussed in Ref.~\cite{Sun2012}.

The tri-block patchy particles do not directly provide additional potential energy penalties to eliminate these floppy modes~\cite{DeGennes1992,Jiang2010}.  Patchy particles are characterized by chemical coatings (\lq\lq patches\rq\rq) that cover finite fractions of their surfaces, and the coated parts of the surface exhibit interactions of very different nature from the uncovered parts.  The interactions between these particles are typically very short-ranged, and thus due to the extended sizes of the patches, these patchy particles often do not have a direct selection of bond-angle directions, as we will discuss in detail in Sec.~\ref{SEC:FEgeneral}.  
The potential energy of pair-wise interactions between these patchy particles are of step-function nature as relative particle-orientations are changed, rather than gradual dependence with a potential energy minimum at certain angle.  

\begin{figure}
	\centering
		\subfigure[]{\includegraphics[width=.23\textwidth]{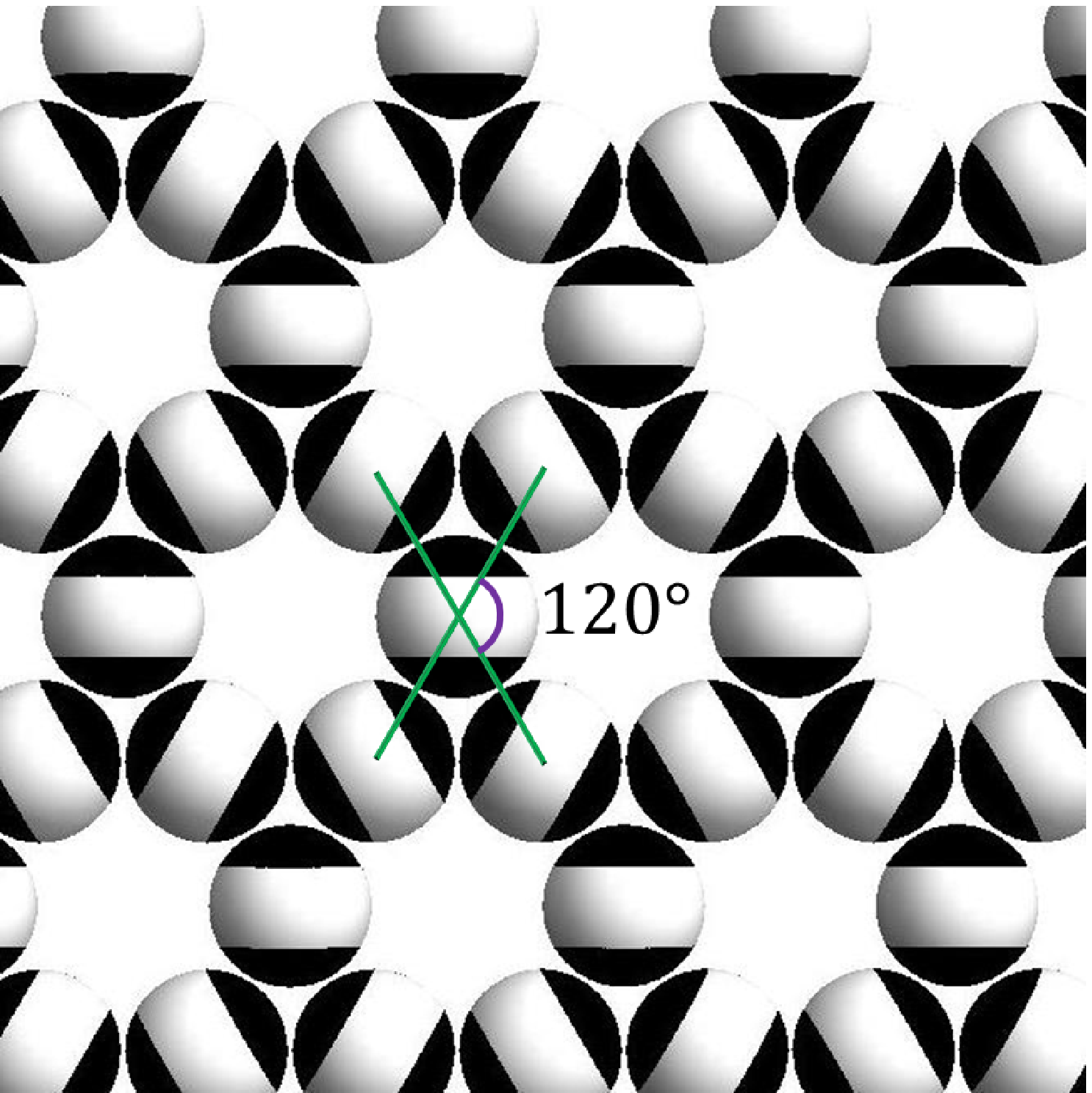}}
		\subfigure[]{\includegraphics[width=.23\textwidth]{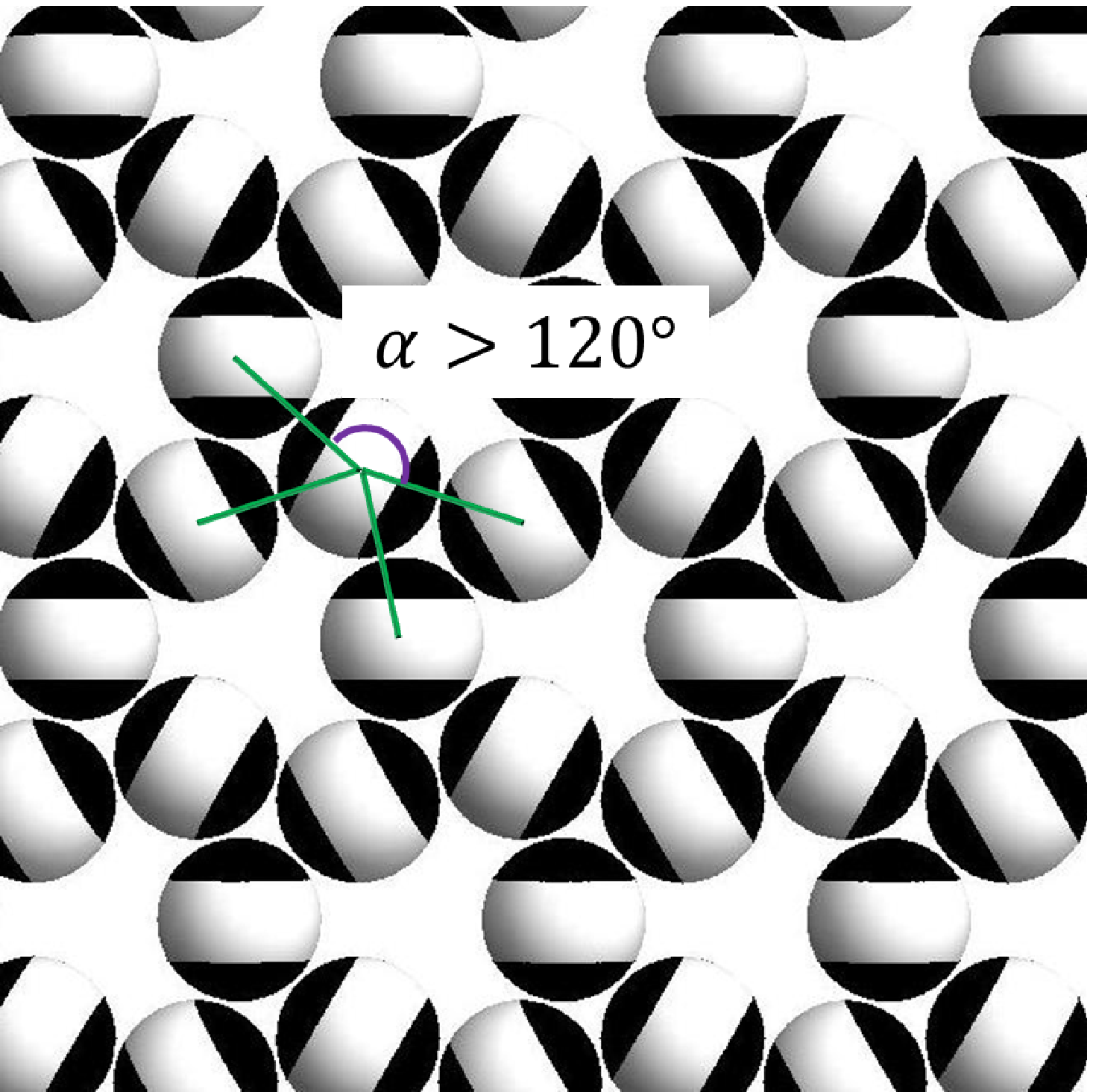}}
		\subfigure[]{\includegraphics[width=.23\textwidth]{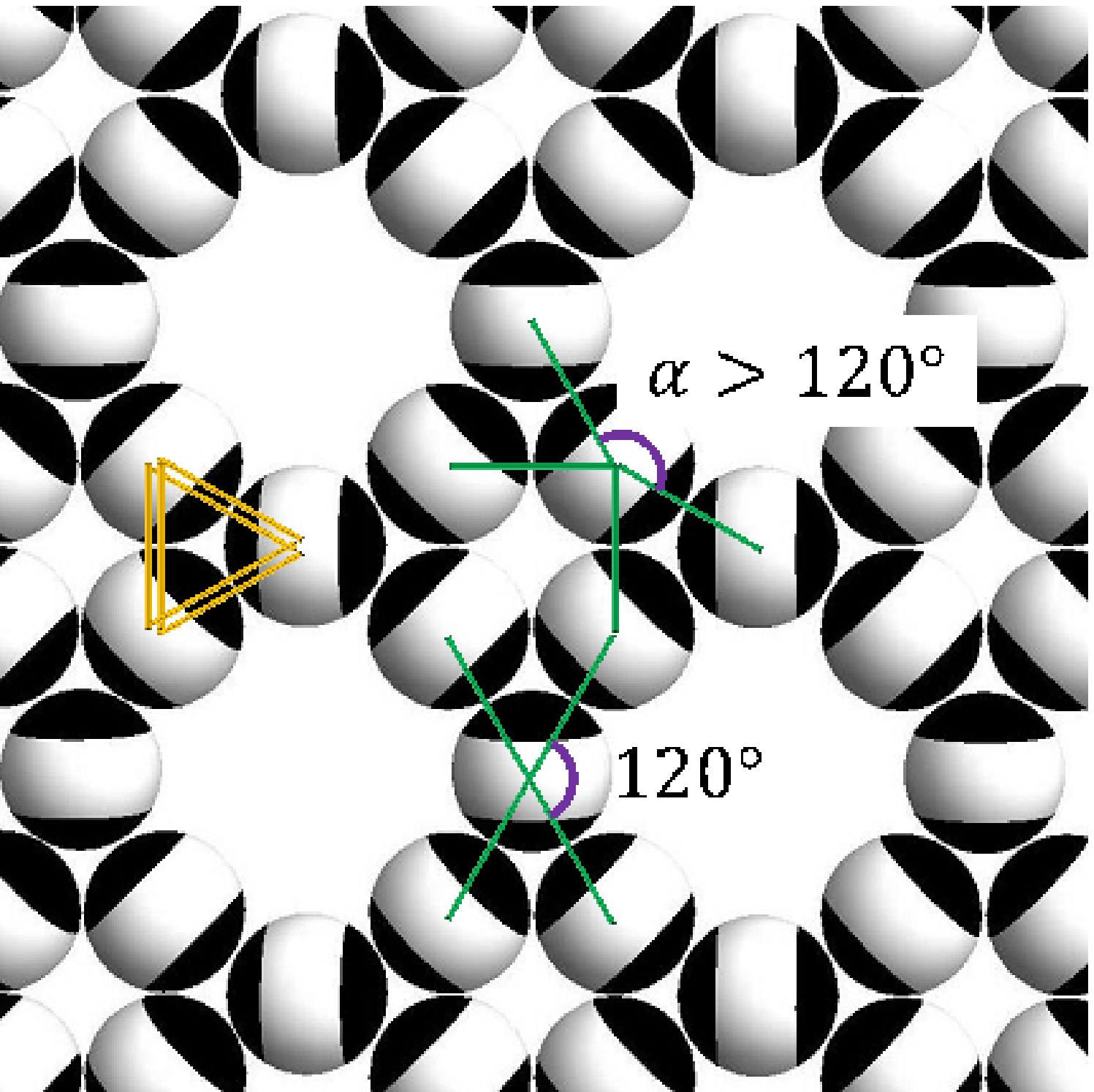}}
		\subfigure[]{\includegraphics[width=.23\textwidth]{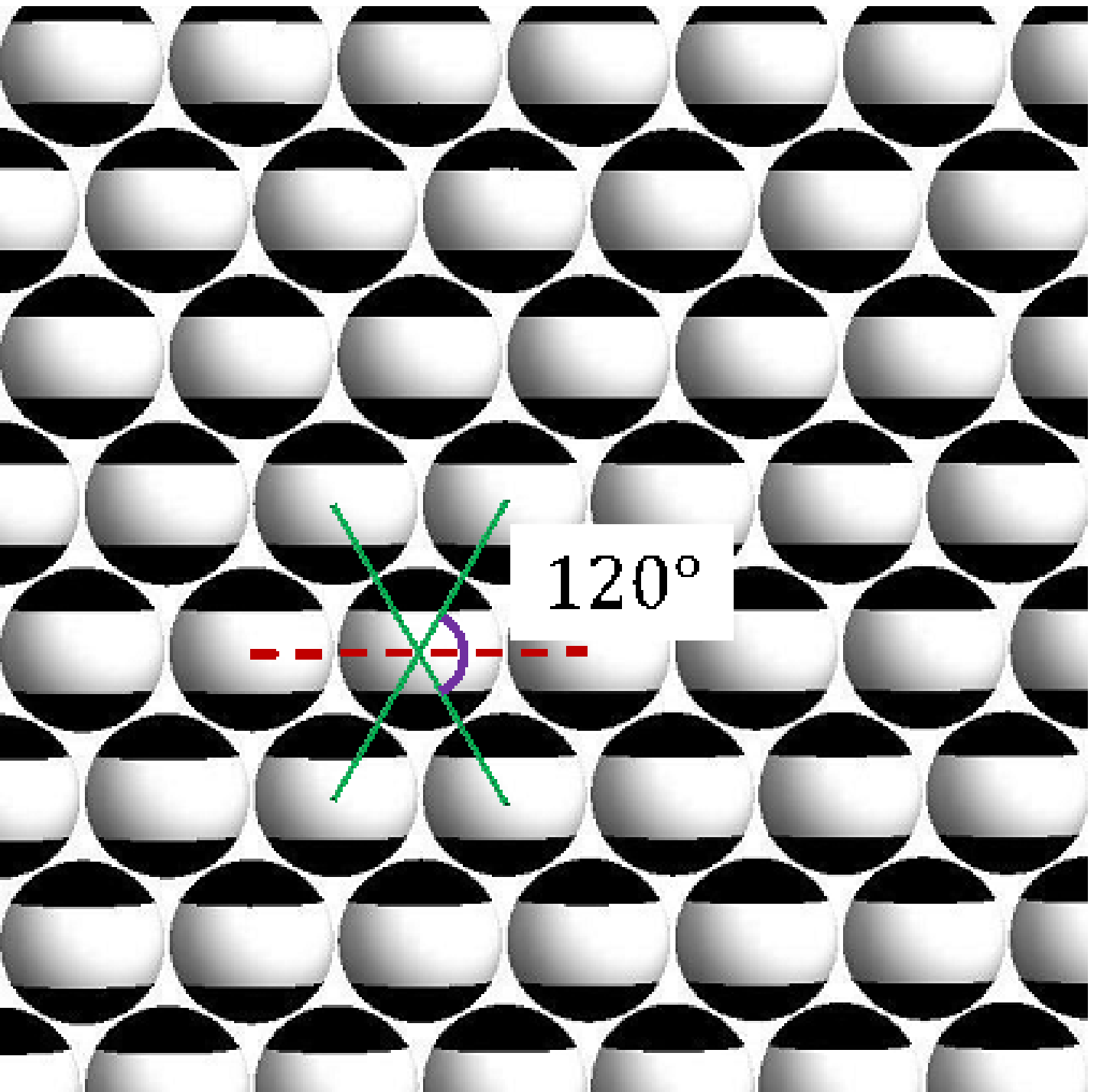}}
	\caption{Examples of possible structures formed by triblock Janus particle of the same potential energy: (a) kagome lattice having bond angle $120^{\circ}$, (b) twisted kagome lattice with bond angle deviating from $120^{\circ}$ (the twisted kagome and the kagome lattice is related by a floppy mode), (c) roman mosaic lattice consists of two types of particles with bond angles $120^{\circ}$ and $150^{\circ}$ ($90^{\circ}$ on the opposing side), and (d) hexagonal lattice with bond angle $120^{\circ}$ but also contacts in non-attractive patch (red dashed lines).  The triangle denoted by orange double lines in (c) marks a corner-sharing-triangle which defines a basic unit in the structure.}
	\label{FIG:IsoEner}
\end{figure}

For the case of tri-block Janus particles discussed in Refs.~\cite{Chen2011,Mao2013}, two particles experience a short-ranged mutual attraction when they are facing each other in their coated attractive patch, and strongly-screened electro-static repulsion (which can be modeled as hard-sphere repulsion) otherwise.  The pair-wise interaction potential changes abruptly at the boundary of the patches, and is almost flat within the patches.  
This flatness of the potential energy as bond-angle changes is the so called \lq\lq degenerate valency\rq\rq~\cite{Chen2011,Mao2013}, and two consequences follow: (i) the floppy modes of the central-force kagome lattice remain zero potential energy in the kagome lattice composed of tri-block patchy particles, and (ii) the kagome lattice is isoenergetic with other lattices which have the same number of contacts per particle in their attractive patches.  This is shown in Fig.~\ref{FIG:IsoEner}.  In addition, we shall call the area of the particle surface that is not coated the \lq\lq non-attractive\rq\rq  patch.

Surprisingly, this flatness in potential energy is lifted by thermal fluctuations via the so called \lq\lq order by disorder\rq\rq  effect~\cite{Villain1980,Henley1987,Chubukov1992,Reimers1993,CastroNeto2006}.  A stable kagome lattice has been observed from the self-assembly of tri-block Janus particles at finite temperature in the experiment discussed in Ref.~\cite{Chen2011}.  In this Paper we discuss a theory describing this effect of how open lattices can be stabilized by entropy.

In Sec.~\ref{SEC:FEgeneral} we set up the equilibrium statistical mechanics for lattices assembled from patchy particles and discuss small fluctuations around the stable reference state.  In Sec~\ref{SEC:Entropy} we discuss the effects of rotational and vibrational entropy of the patchy particles and how they contribute to the stability of open lattices.  In Sec~\ref{SEC:Results} we summarize the results of our theory applied on the system of triblock Janus particles.  In Sec~\ref{SEC:Discussion} we discuss the agreement of our theory with experiment, possible improvements, relations and applications to other systems.

%%%%%%%%%%%%%%%%%%%%%%%%%%%%%%%%%%%%%%
%%%%%%%%%%%%%%%%%%%%%%%%%%%%%%%%%%%%%%
%%%%%%%%%%%%%%%%%%%%%%%%%%%%%%%%%%%%%%
\section{Free energy of open lattices assembled from patchy particles: general formulation}\label{SEC:FEgeneral}
To model the equilibrium statistical mechanics of a colloidal open lattice, we start from the generic form of partition function applicable to anisotropic particles,
\begin{align}
	Z=\int e^{-\frac{H(\{\vr_i,\ori_i\})}{k_B T}} \prod _j d\vr _j d\ori_j ,
\end{align}
where $H$ is the Hamiltonian, $\vr_i,\ori_i$ are the position and orientation of the particle $i$, $k_B$ is the Boltzmann's constant, and $T$ is the temperature.  
For a three-dimensional particle, the orientation $\ori_i$ represent the information of the polar and azimuth angle of the particle.  If the particle does not have symmetry for rotations around its own axes $\ori_i$ an additional variable is needed to fully describe its orientation.  For simplicity we just consider particles with this symmetry and thus $\ori_i$ is sufficient.

\begin{figure}
	\centering
		\includegraphics[width=.4\textwidth]{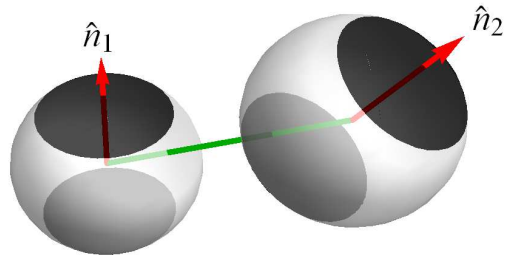}
	\caption{Pair-wise interaction between two patchy particles.  The green line denotes the vector connecting the centers of the two particles, $\vr_j - \vr_i$.  The unit vector $\uve_{ij}=(\vr_j - \vr_i)/\vert \vr_j - \vr_i\vert$.  The red arrows denote the orientations of the two particles, $\ori_i$ and $\ori_j$.  The surface elements crossed by the green line on the two particles dominate the interaction in Eq.~\eqref{EQ:Vtwo} at short range.}
	\label{FIG:twobody}
\end{figure}

Now we specialize our discussion to the case spherical shaped colloidal particles with inhomogeneous surface coatings (patches).  The triblock Janus particles used in the experiment in Ref.~\cite{Chen2011} is one example of such particles.  For these particles the Hamiltonian of the system consists pair-wise potential energy of the form
\begin{align}\label{EQ:Vtwo}
	\mathcal{V} (\vr_i,\ori_i,\vr_j,\ori_j) =& \int d\Omega_i d\Omega_j 
	\tilde{v} \Big(\ori_i,\ori_j,\uvm_{i},\uvm_{j}, \nonumber\\
	&\big\vert (\vr_i+a \uvm_{i}/2)-(\vr_j+a \uvm_{j}/2)\big\vert
	\Big),
\end{align}
which comes from the interactions $\tilde{v}$ between the surface elements of the two particles, integrated over the solid angles of both of them.  In this expression $a$ is the diameter of the particle, $\uvm_{i}$ and $\uvm_{j}$ are unit vectors pointing in the directions of the solid angles $\Omega_i$ and $\Omega_j$ respectively, and thus $\big\vert (\vr_i+a \uvm_{i}/2)-(\vr_j+a \uvm_{j}/2)\big\vert$ is the distance between the two surface elements.  Figure~\ref{FIG:twobody} illustrate such a pair-wise interaction.  If the range of interactions between the surface elements of the two particles is much shorter than the particle size, and the two particles are sufficiently close, the surface elements crossed by the vector pointing from the center of particle $i$ to $j$ dominates the integral and we can write
\begin{align}
	\mathcal{V} (\vr_i,\ori_i,\vr_j,\ori_j) \simeq & \tilde{v}\Big(\ori_i,\ori_j,\uve_{ij},-\uve_{ij}, \nonumber\\
	&\big\vert (\vr_i+a \uve_{ij}/2)-(\vr_j-a \uve_{ij}/2)\big\vert
	\Big),
\end{align}
where $\uve_{ij}$ is the unit vector pointing from particle $i$ to $j$.

We further specialize to patchy particles that exhibit short-ranged attractions when they face each other in their attractive patches and hard-sphere repulsion otherwise, and consider the free energy of equilibrium colloidal lattices assembled from these particles.   
For these patchy particles it is often practical to assume that (i) the depth of the attractive well is sufficiently greater than thermal energy (in the case of Ref.~\cite{Chen2011} the depth is approximately $7k_B T$), and (ii) the boundaries of the attractive patch is sharply defined, i.e., the interactions are attractive when two particles are facing each other in their attractive patch, and rapidly changes into hard-sphere repulsion otherwise.  

Thus for our consideration of the free energy of equilibrium colloidal lattices, for which it is sufficient to consider only small fluctuations away from the stable state, we can make further simplifications of the problem.  For a given lattice structure, we can identify all contacts between neighboring particles in their attractive patch, which we shall call \lq\lq attractive bonds\rq\rq , and in the calculation of the partition function only include those states in which these attractive bonds are not broken.  All states with broken attractive bonds have a very small Boltzmann factor and can thus be ignored.  For these attractive bonds, the potential energy can then be written into the form
\begin{align}\label{EQ:Vvw}
	\mathcal{V} (\vr_i,\ori_i,\vr_j,\ori_j) = v_a(\vert \vr_i-\vr_j\vert) w(\hat{e}_{ij},\ori_i)
	w(-\hat{e}_{ij},\ori_j)
\end{align}
where the factors
\begin{align}
	w(\hat{e}_{ij},\ori_i) = \left\{ \begin{array}{ll}
		1 & \textrm{if $\hat{e}_{ij}$ passes through the}\\
		& \textrm{ attractive patch of particle $i$} \\
		\infty & \textrm{otherwise}
	\end{array} \right.
\end{align} 
and the central-force part of the interaction, $v_a(\vert \vr_i-\vr_j\vert)$ is characterized by a hard-core repulsion, a short range attraction, and vanishing interaction when the distance between the centers of the particles is greater than the radius plus the range of attraction.  

There can also be contacts in the non-attractive surfaces of the particles in the stable state (we shall call them \lq\lq non-attractive bonds\rq\rq), as we shall discuss in the case of the hexagonal lattice.  These contacts are simply described by isotropic hard-sphere repulsions, given the fact that the stable state we start from is a local minimum in potential energy and these contacts will not turn into an attractive contact unless other attractive bonds are broken.  Thus the potential of these non-attractive bonds is simply
\begin{align}
	\mathcal{V} (\vr_i,\ori_i,\vr_j,\ori_j) = v_r(\vert \vr_i-\vr_j\vert) ,
\end{align}
where $v_r$ describes a very short-ranged repulsion.

\begin{figure}
	\centering
		\subfigure[]{\includegraphics[width=.18\textwidth]{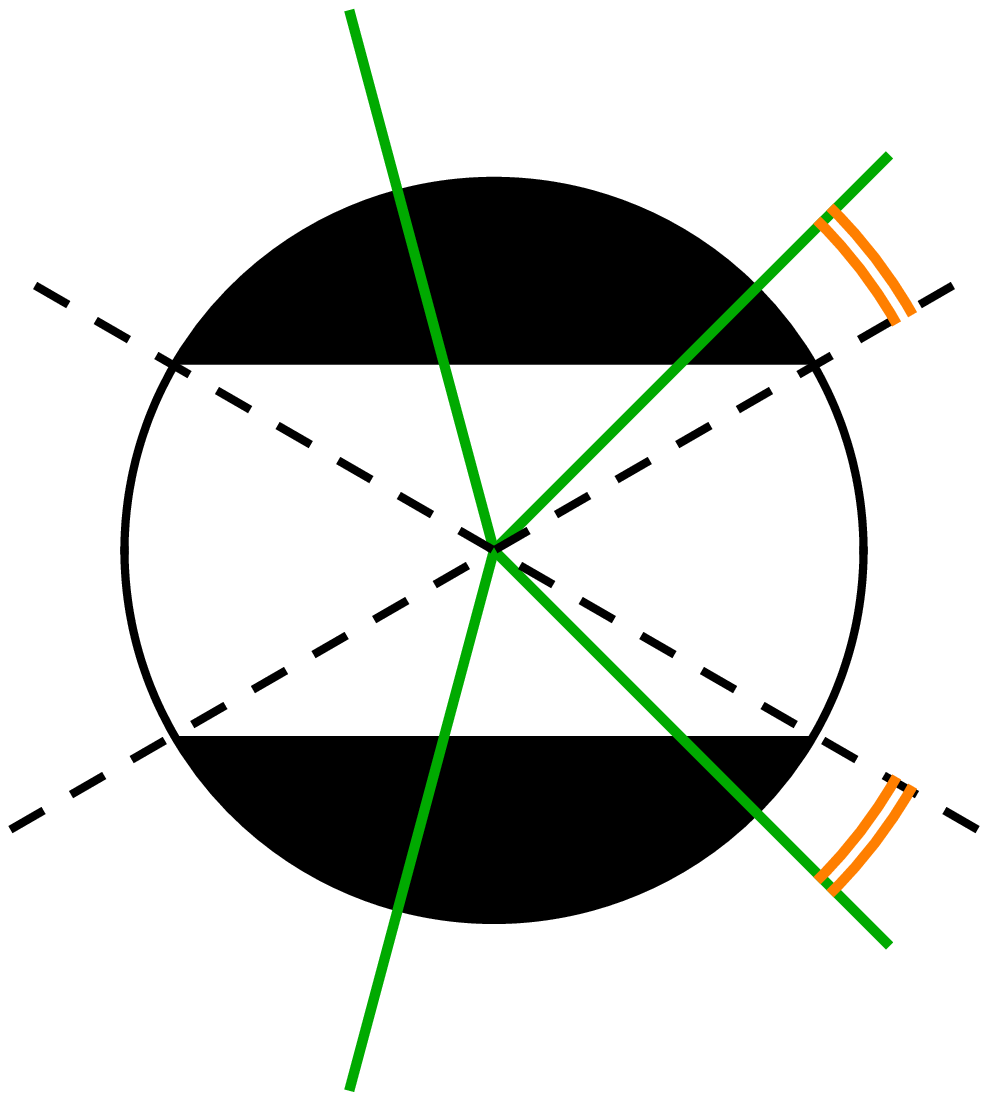}} \quad
		\subfigure[]{\includegraphics[width=.18\textwidth]{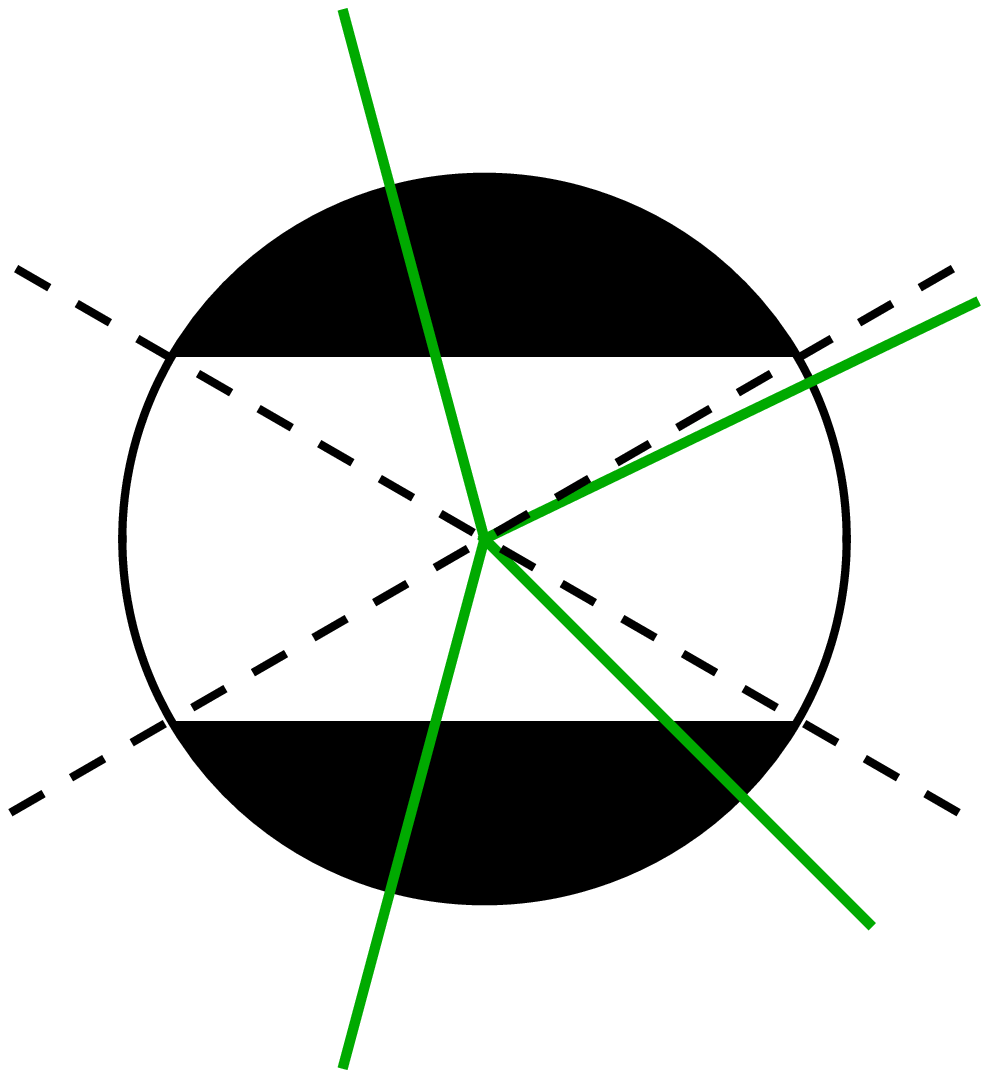}}
	\caption{(a) An example configuration that gives $\Psi_j\big( \{\vr_i\},\ori_j \big) =1$.  In this configuration all bonds (green solid lines) connecting particle $j$ go through its attractive patch (black area).  The orange double arcs represent the allowed microscopic orientational states of particle $j$ given the bonds.  (b) An example configuration that gives $\Psi_j\big( \{\vr_i\},\ori_j \big) =0$.  One bond goes through the non-attractive patch.  The black dashed lines mark the boundary of the attractive patches.  Rules for three dimensional configurations follow similarly.  }
	\label{FIG:Psi}
\end{figure}

With the form of potential energy~\eqref{EQ:Vvw} for the attractive bonds, we can decouple the orientational degrees of freedom of different particles as follows.  Because $w\to\infty$ if any of the particle has an orientation that one or more attractive bond is broken, and thus the corresponding Boltzmann factor vanishes in the summation of the partition function.  We have
\begin{align}\label{EQ:Zdecouple}
	Z&=\int e^{-\frac{1}{k_B T}\sum_{\langle mn \rangle}\mathcal{V} (\vr_m,\ori_m,\vr_n,\ori_n)} \prod _j d\vr _j d\ori_j 
	\nonumber\\
	&= \int e^{-\frac{\HCF( \{\vr_i\} )}{k_B T}} \prod _j \Psi_j\big( \{\vr_i\},\ori_j \big) d\vr _j d\ori_j 
\end{align}
where the sum is over nearest bonds $\langle mn \rangle$.  The central-force Hamiltonian is given by
\begin{align}
	\HCF\big( \{\vr_i\} \big) = \sum_{\langle mn \rangle_a}v_a(\vert \vr_m - \vr_n \vert) +
	\sum_{\langle mn \rangle_r}v_r(\vert \vr_m - \vr_n \vert) ,
\end{align}
where $\langle mn \rangle_a$ represent attractive bonds and $\langle mn \rangle_r$ represent non-attractive bonds. 
The factor $\Psi_j\big( \{\vr_i\},\ori_j \big) =1$ if all of the attractive bonds of particle $j$ remain in its attractive patch given its orientation, and 0 if any attractive bonds of $j$ is broken.  A two-dimensional example of $\Psi_j$ function is shown in Fig.~\ref{FIG:Psi}.  It is clear that besides the positions of particles, this $\Psi$ factor \emph{only depends on the orientation of one particle, regardless of the orientation of all other particles}.  Our assumption of small fluctuations near the equilibrium state allowed this decoupling.  Therefore, from Eq.~\eqref{EQ:Zdecouple} we can integrate out the orientational degrees of freedom of each particles individually.  The result from each of these integration
\begin{align}\label{EQ:MS}
	\MS_j = \int \Psi_j \big( \{\vr_i\},\ori_j \big) d\ori_j  ,
\end{align}
is proportional to the number of microscopic orientational states allowed for particle $j$ given the positions of this particle and its neighbors.  It is appropriate to use the concept of microscopic orientational states here, because with fixed positions, the allowed rotation of one particle can be described by micro-canonical ensemble, since all allowed orientations are of the same energy.  
The result of this integral can be written in terms of the rotational entropy
\begin{align}
	s_j = k_B \ln \MS_j ,
\end{align}
which adds up with the central-force part of the potential energy and lead to an effective Hamiltonian that depends on particle positions only
\begin{align}\label{EQ:HEF}
	\HEF = \HCF - T \sum_j s_j .
\end{align}
The partition function is then
\begin{align}\label{EQ:ZEF}
	Z=\int e^{-\frac{\HEF( \{\vr_i\} )}{k_B T}} \prod _j  d\vr _j . 
\end{align}
We shall discuss this calculation of the rotational entropy in detail for the case of triblock Janus particles below.

%%%%%%%%%%%%%%%%%%%%%%%%%%%%%%%%%%%%%%
%%%%%%%%%%%%%%%%%%%%%%%%%%%%%%%%%%%%%%
%%%%%%%%%%%%%%%%%%%%%%%%%%%%%%%%%%%%
\section{Effects of entropy: the example of triblock Janus particle}\label{SEC:Entropy}
In this section, we study the effects of entropy using the formulation discussed in Sec.~\ref{SEC:FEgeneral}.  We use the triblock Janus particles system as an example to show how entropy provides mechanical stability and induces differences in the free energy of open and close-packed lattices which have the same potential energy.

We will discuss two different type of triblock Janus particles, corresponding to two recent experiments.  The first type of triblock Janus particles have attractive patches that are elongated in the plane of the lattice~\cite{Chen2011}.  As a result for the lattices we discuss in this Paper, the kagome and the hexagonal lattices, in which each particle has $4$ attractive bonds, the orientational fluctuations of the particle is confined to two dimensions in the calculation of partition function~\eqref{EQ:Zdecouple}.
The second type of triblock Janus particles have circular shaped attractive patches, and the allowed orientational fluctuations are three dimensional~\cite{Chen2011b}.  We will discuss both cases in the following.

%%%%%%%%%%%%%%%%%%%%%%%%%%%%%%%%%%%%%%
%%%%%%%%%%%%%%%%%%%%%%%%%%%%%%%%%%%%%%
\subsection{Rotational entropy}
%%%%%%%%%%%%%%%%%%%%%%%%%%%%%%%%%%%%%%
\subsubsection{Triblock Janus particles with elongated patches}
These triblock particles can be considered as two-dimensional disks that assemble into two-dimensional lattices, and their orientations are characterized by only the polar angle $\theta_i$.  

\begin{figure}
	\centering
		\includegraphics[width=.2\textwidth]{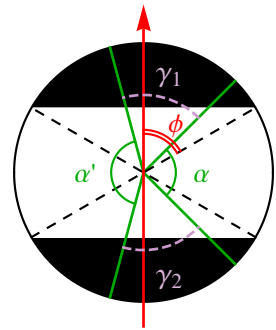}
	\caption{A triblock Janus particle with rotations confined in two-dimensional space, due to the elongated patch shape (not shown in this figure).  The patch size $\hps$ and the bond angles $\alpha,\alpha',\g_1,\g_2$ are also shown.  The $4$ green solid lines represent the $4$ attractive patches of the particle, and the red arrow represent the orientation of the particle.  The black dashed lines represent the boundaries of the attractive patches.}
	\label{FIG:triblock1}
\end{figure}

As shown in Fig.~\ref{FIG:triblock1}, to keep all attractive bonds of particle $i$ in its attractive patch, its orientation $\theta_i$ is confined by the bond angles, $\alpha,\alpha',\g_1,\g_2$ (here and in the following discussion in this subsection we drop for subscript $i$ for convenience), between the $4$ attractive bonds.  It is clear that there are only $3$ independent variables because
\begin{align}
	\alpha +\alpha' +\g_1 +\g_2 =2\pi .
\end{align}
The values of these angles in the stable state are
\begin{align}
	\alpha=\alpha'=\frac{2\pi}{3}, \quad \g_1=\g_2=\frac{\pi}{3} ,
\end{align}
and they deviate from these values at finite temperature.  

As we found by comparing to experimental data, in lattices assembled from triblock Janus particles, the stiffness against bond-angle fluctuations is much smaller than the stiffness against bond-length fluctuations~\cite{Mao2013}.  Because the two pairs of nearest neighbors of the $4$ attractive bonds one triblock Janus particle (in Fig.~\ref{FIG:triblock1}, the top pair and the bottom pair) are themselves nearest-neighbor pairs, it is straightforward to realize that the fluctuations in the angles $\g_1$ and $\g_2$ are much smaller than those of the angles $\alpha$ and $\alpha'$.  Therefore, it is reasonable to make the simplifying assumption of ignoring the fluctuations in $\g_1$ and $\g_2$, so that
\begin{align}
	\g_1 = \g_2 =\frac{\pi}{3} .
\end{align}
Then we have
\begin{align}
	\alpha +\alpha' = \frac{4\pi}{3},
\end{align}
and there is only one bond-angle variable $\alpha$.  We also show the calculation of rotational entropy with fluctuations in $\g_1, \g_2$ allowed in Sec.~\ref{SEC:GAMMA}.  The calculation is considerably more complicated but the result is very close of that of the current simplified model.

Given $\alpha$ we consider the allowed orientations $\theta$ of the particle.  Choosing the polar coordinate so that at $\theta=0$ the \lq\lq north pole\rq\rq  of the particle is aligned with the axes of reflection symmetry of the $4$ attractive bonds, as shown in Fig.~\ref{FIG:triblock1}.  The total allowed rotational microscopic states of the particle, as introduced in Eq.~\eqref{EQ:MS}, is thus proportional to
\begin{align}
	\MS =& \int_{-\pi}^{\pi} d\theta \Psi\big( \{\vr_i\},\theta \big) \nonumber\\
	=& \int_{-\pi}^{\pi} d\theta \,\Theta \left(\theta+\hps-\frac{\pi-\alpha}{2} \right) \nonumber\\
	& \times \Theta \left(\frac{\pi+\alpha}{2} -(\theta+\pi-\hps)\right)
\end{align}
where $\hps$ is the angle characterizing the half patch size as shown in Fig.~\ref{FIG:triblock1}, and without loss of generality we assumed $\alpha<\alpha'$.  We used the Heaviside step function, $\Theta$, to enforce our condition of the $\Psi$ functions,  that the allowed orientations $\theta$ need to keep all attractive bonds in the attractive patches of this particle.  Introducing the bond angle fluctuation variable
\begin{align}
	\Delta \alpha = \alpha - \frac{2\pi}{3},
\end{align}
we have
\begin{align}\label{EQ:MSS}
	\MS = 2\left(\hps -\frac{\pi}{6}\right) - \vert \Delta \alpha \vert
\end{align}
where the absolute value operation on $\Delta \alpha$ comes from releasing the assumption that $\alpha<\alpha'$.  This result leads to the rotational entropy of the particle given bond angle change $\Delta\alpha$ ,
\begin{align}\label{EQ:sElon}
	s=k_B \ln \left\lbrack 2\left(\hps -\frac{\pi}{6}\right) - \vert \Delta \alpha \vert \right\rbrack .
\end{align}
This rotational entropy is shown together with that of the particles of circular patches in Fig.~\ref{FIG:RotEnt}.   
For any given set of particle positions $\{\vr_i\}$ all the bond angles are given and we can calculate the rotational entropy $s_i$ for each of the particles.

%%%%%%%%%%%%%%%%%%%%%%%%%%%%%%%%%%%%%%
\subsubsection{Triblock Janus particles with circular patches}
In contrast to our discussion on Janus particles with elongated patches, particles with circular patches need to be described using three-dimensional rotations with polar angle $\theta$ and azimuth angle $\psi$.

\begin{figure}
	\centering
		\includegraphics[width=.2\textwidth]{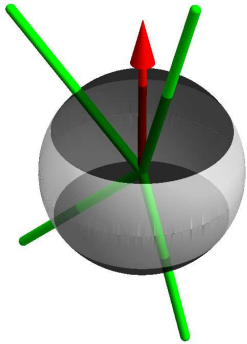}
	\caption{$3D$ rotation of the sphere.  Green lines denote the bonds, and the red arrow denote the north pole of the particle.}
	\label{FIG:Rot3D}
\end{figure}
For small fluctuations around the stable state, we still require that the $4$ attractive bonds remain in the attractive patch.  Given that the $4$ attractive bonds are in the same plane (for two-dimensional kagome and hexagonal lattices), and $\g_1=\g_2=\pi/3$, we can write down the unit vectors pointing in the directions of these $4$ attractive bonds
\begin{align}
	\vec{M}_1 &= \left\{ \sin\left(\frac{\pi}{2} - \frac{\alpha}{2}\right),  0, 
		\cos\left(\frac{\pi}{2} - \frac{\alpha}{2}\right) \right\} , \nonumber\\
	\vec{M}_2 &= \left\{ \sin\left(\frac{\pi}{2} - \frac{\alpha}{2}\right) ,0,
		-\cos\left(\frac{\pi}{2}- \frac{\alpha}{2}\right) \right\} , \nonumber\\
	\vec{M}_3 &= \left\{ -\sin\left(\frac{\pi}{2} - \frac{\alpha'}{2}\right),  0, 
		\cos\left(\frac{\pi}{2} - \frac{\alpha'}{2}\right) \right\} , \nonumber\\
	\vec{M}_4 &= \left\{ -\sin\left(\frac{\pi}{2} - \frac{\alpha'}{2}\right) ,0,
		-\cos\left(\frac{\pi}{2}+ \frac{\alpha'}{2}\right) \right\} , 
\end{align}
where $\alpha$ and $\alpha'$ are the bond angles on the left and the right and they are related through
\begin{align}
	\alpha+\alpha'=4\pi/3 .
\end{align}
We then keep these $4$ bonds fixed in space and let the particle rotate around its center, and calculate the rotational microscopic states of this particle given that all $4$ attractive bonds are kept in its attractive patches.  The direction of the \lq\lq north pole\rq\rq\ of the particle
\begin{align}
	\vec{N} = \left\{ \sin\theta \cos\psi,  \sin\theta \sin\psi , 
		\cos\theta \right\} ,
\end{align}
fully describes the orientation of this particle, because the patches of the triblock Janus particles are symmetric with respect to rotations around its north pole.  If the patches are of more complicated patterns, more variables will be needed to describe it.

\begin{figure}[h]
	\centering
		\includegraphics[width=.45\textwidth]{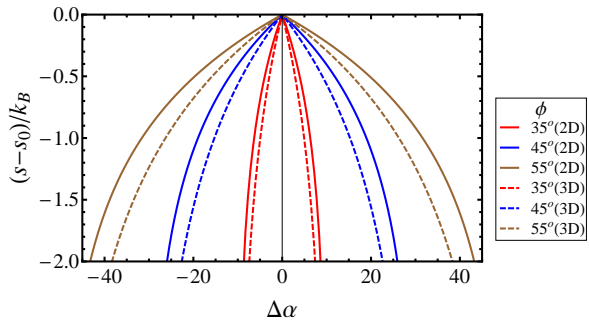}
	\caption{Rotational entropy of the triblock Janus particles with elongated patch (and thus $2D$ rotations), denoted by solid lines, and circular patch (and thus $3D$ rotations), denoted by dashed lines, as a function of the bond angle difference $\Delta\alpha$ at different patch size $\hps$ as listed in the legend.  The curves from outer to inner represent the change of rotational entropy $s-s_0$ in cases of $\hps=55^{\circ},45^{\circ},35^{\circ}$ respectively, where $s_0$ denote the rotational entropy when $\Delta\alpha=0$.}
	\label{FIG:RotEnt}
\end{figure}

Again, without losing any generality we assume that $\alpha<\alpha'$.  The condition that all four bonds remain in the attractive patch is then
\begin{align}
	\vec{N}\cdot\vec{M}_1 \ge \cos\hoa , \nonumber\\
	-\vec{N}\cdot\vec{M}_2 \ge \cos\hoa  ,
\end{align}
which simplify into
\begin{align}\label{EQ:EQL}
	\cos\theta \sin\frac{\alpha}{2}  \pm  \sin\theta\cos\psi\cos\frac{\alpha}{2}\ge \cos\hoa  .
\end{align}
The total number of rotational microscopic states is thus proportional to
\begin{align}
	\MS = &\int_{0}^{\pi} d\theta \int_{-\pi}^{\pi}d\psi \nonumber\\
	&\times\Theta \left( \cos\theta \sin\frac{\alpha}{2}  
	+  \sin\theta\cos\psi\cos\frac{\alpha}{2}- \cos\hoa\right) \nonumber\\
	&\times \Theta \left( \cos\theta \sin\frac{\alpha}{2}  -  \sin\theta\cos\psi\cos\frac{\alpha}{2}- \cos\hoa\right).
\end{align}
Equation~\eqref{EQ:EQL} indicates two critical value for $\theta$
\begin{align}
	\theta_1 &= \frac{\alpha}{2}+\phi-\frac{\pi}{2} , \nonumber\\
	\theta_2 &= \arccos  \left( \frac{\cos\phi}{\sin\frac{\alpha}{2}}\right) ,
\end{align}
that for $0<\theta<\theta_1$ the equality~\eqref{EQ:EQL} is satisfied for all values of $\psi$, and for $\theta_1<\theta<\theta_2$ the equality~\eqref{EQ:EQL} is only satisfied for some values of $\psi$, in this case, around $\psi=\pm\pi/2$.  Therefore the number of rotational states in the three-dimensional rotation case is proportional to 
\begin{align}
	\Omega = \Omega^{(1)}+\Omega^{(2)}
\end{align}
with
\begin{align}
	\Omega^{(1)} &= 2\pi ( 1-\cos\theta_1 ) \nonumber\\
	\Omega^{(2)} &= \int _{\theta_1}^{\theta_2} d\theta \sin\theta \,\, 4 \Big\lbrack
		\frac{\pi}{2} - \arccos\left(\frac{\cos\hoa 
		- \cos\theta \sin\frac{\alpha}{2}}{\sin\theta\cos\frac{\alpha}{2}}\right)
	\Big\rbrack .
\end{align}
We calculated this number of rotational states and plot the results in Fig.~\ref{FIG:RotEnt} in the paper.  It has no qualitative difference from the $2D$ case.

%%%%%%%%%%%%%%%%%%%%%%%%%%%%%%%%%%%%%%
\subsubsection{Correction to rotational entropy allowing bond length fluctuations}\label{SEC:GAMMA}
A more realistic description of the rotational entropy should allow the fluctuations in bond length.  In fact in Sec.~\ref{SEC:VE} we shall discuss the contribution of these vibrational modes to the free energy.  These bond length fluctuations lead to fluctuations in the angles $\g_1$ and $\g_2$, which we took to be fixed value $\pi/3$ in our previous discussions.  For convenience we define the deviations of the angles
\begin{align}
	&\alpha=2\pi/3+\Delta \alpha , \quad
	\alpha'=2\pi/3+\Delta \alpha'  , \nonumber\\
	&\g_1=\pi/3+\Delta \g_1 ,\quad
	\g_2=\pi/3+\Delta \g_2 ,
\end{align}
and they satisfy the equation
\begin{eqnarray}
	\Delta \alpha+\Delta \alpha'+\Delta \g_1+\Delta \g_2=0.
\end{eqnarray}
Thus we just choose $\Delta \alpha,\Delta \g_1, \Delta \g_2$ to be the independent variables.  We now calculate the rotational entropy in the two-dimensional case.  The case of three-dimensional rotations follow similarly but involve more complicated derivations.

\begin{figure}
	\centering
		\subfigure[]{\includegraphics[width=.15\textwidth]{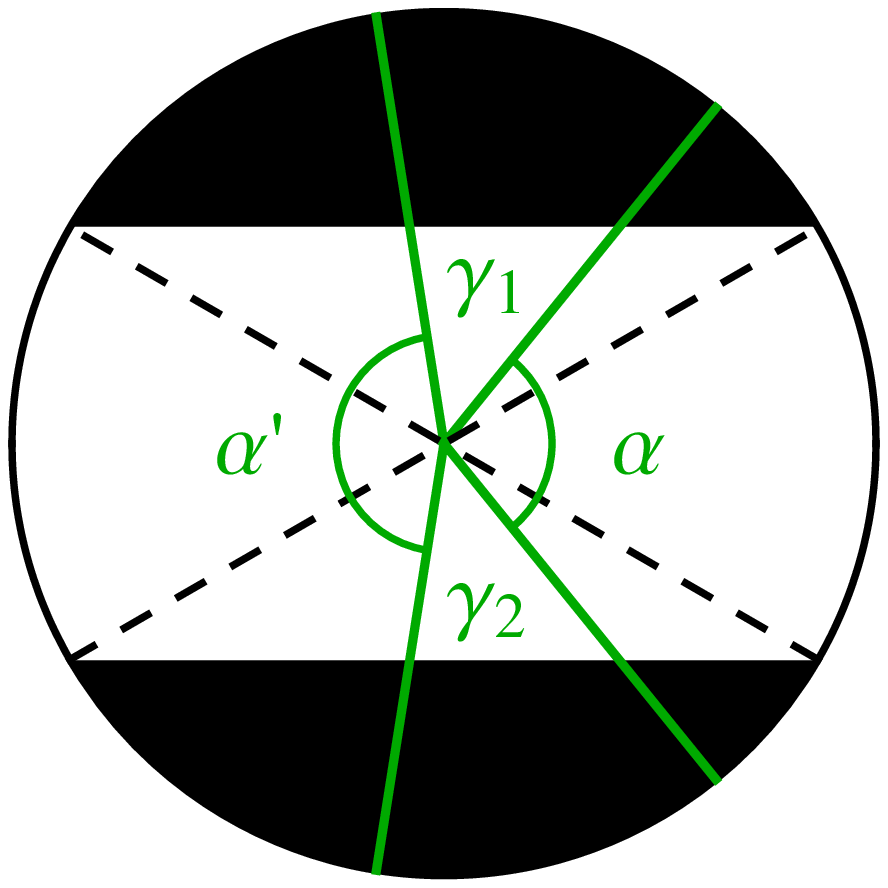}}
		\subfigure[]{\includegraphics[width=.15\textwidth]{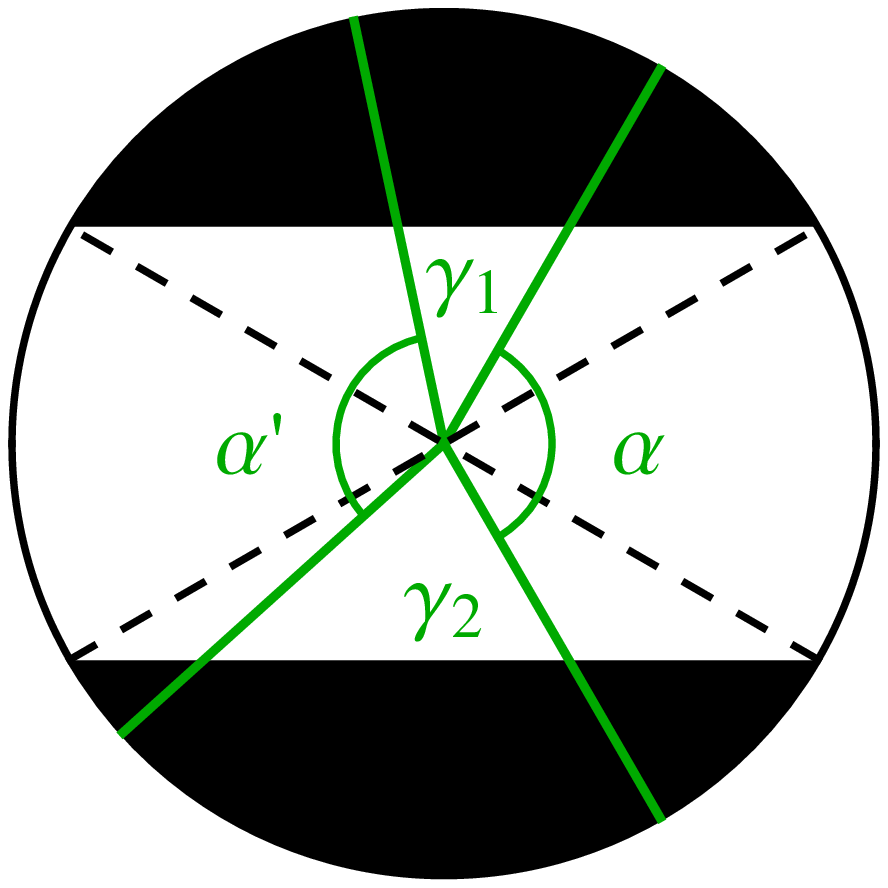}}\\
		\subfigure[]{\includegraphics[width=.15\textwidth]{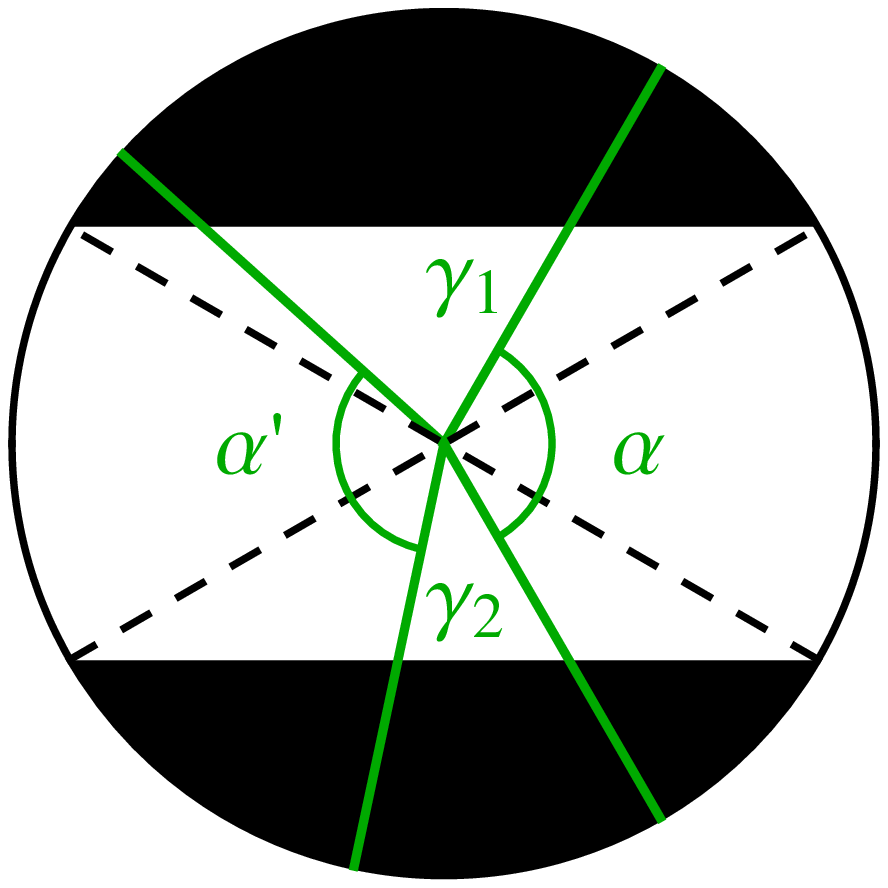}}
		\subfigure[]{\includegraphics[width=.15\textwidth]{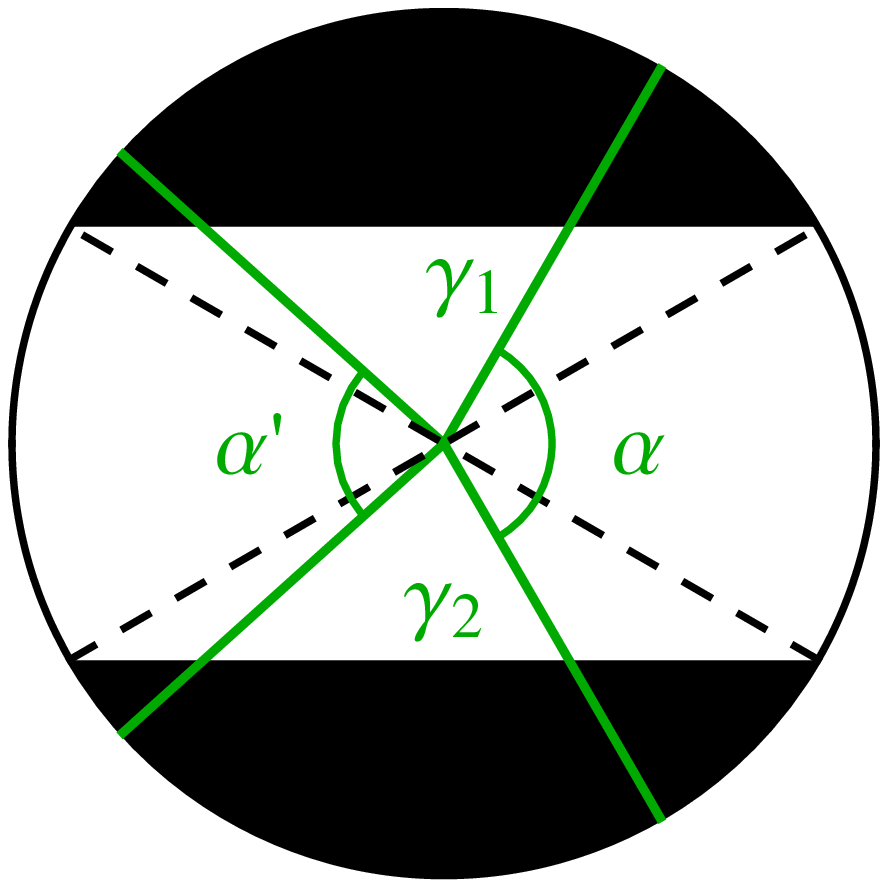}}
	\caption{Bond angle configurations at given $\g_1$ and $\g_2$ in the $4$ situations in Eq.~\eqref{EQ:REC}.  (a) $\Delta \alpha+\Delta\g_1<0$ and $\Delta \alpha+\Delta\g_2<0$.  (b) $\Delta \alpha+\Delta\g_1<0$ and $\Delta \alpha+\Delta\g_2>0$.  (c) $\Delta \alpha+\Delta\g_1>0$ and $\Delta \alpha+\Delta\g_2<0$.  (d) $\Delta \alpha+\Delta\g_1>0$ and $\Delta \alpha+\Delta\g_2>0$.  }
	\label{FIG:ANGLE}
\end{figure}

Similarly, to calculate the number of rotational microscopic states, we that all the $4$ bonds stay in the attractive patch.  We arrive at the following form for the rotational entropy
\begin{widetext}
\begin{eqnarray}\label{EQ:REC}
	\Omega = \left\{
	\begin{array}{ll}
	2(\hoa-\pi/6)+\Delta \alpha 
	& \textrm{if $\Delta \alpha+\Delta\g_1<0$ and $\Delta \alpha+\Delta\g_2<0$} ,\\
	2(\hoa-\pi/6)-\Delta \g_2 
	& \textrm{if $\Delta \alpha+\Delta\g_1<0$ and $\Delta \alpha+\Delta\g_2>0$} ,\\
	2(\hoa-\pi/6)-\Delta \g_1 
	& \textrm{if $\Delta \alpha+\Delta\g_1>0$ and $\Delta \alpha+\Delta\g_2<0$} ,\\
	2(\hoa-\pi/6)-\Delta \g_2-\Delta \g_1-\Delta \alpha
		& \textrm{if $\Delta \alpha+\Delta\g_1>0$ and $\Delta \alpha+\Delta\g_2>0$,}
	\end{array} \right .
\end{eqnarray}
\end{widetext}
for the range of $-2(\hoa-\pi/6)<\Delta \alpha < 2(\hoa-\pi/6)-\g_1-\g_2$ (the number of rotational states vanishes for $\Delta \alpha$ outside this range).  It can be easily understood as follows.  In the case of $\Delta \alpha+\Delta\g_1<0$ and $\Delta \alpha+\Delta\g_2<0$, it is the two bonds on the right that determine the allowed rotations.  In the case of $\Delta \alpha+\Delta\g_1<0$ and $\Delta \alpha+\Delta\g_2>0$, it is the two bonds on the bottom that determine the allowed rotations.   In the case of $\Delta \alpha+\Delta\g_1>0$ and $\Delta \alpha+\Delta\g_2<0$, it is the two bonds on the top that determine the allowed rotations.   In the case of $\Delta \alpha+\Delta\g_1>0$ and $\Delta \alpha+\Delta\g_2>0$, it is the two bonds on the left that determine the allowed rotations.  These different cases are depicted in Fig.~\ref{FIG:ANGLE}.

The deviations $\g_1$ and $\g_2$ are limited by the central-force interactions between nearest neighbors and are thus small.  We can then assume that $\vert \g_1\vert, \vert \g_2\vert \ll 2(\hoa-\pi/6)$ for the range of parameters we are interested in.

%%%%%%%%%%%%%%%%%%%%%%%%%%%%%%%%%%%%%%
%%%%%%%%%%%%%%%%%%%%%%%%%%%%%%%%%%%%%%
\subsection{Vibrational entropy}\label{SEC:VE}
In this section we discuss the next step: integrating out the positional degrees of freedom in the partition function and arrive at the free energy for given lattices.  We shall find that although open and close-packed lattices can have the same potential energy, due to their difference in vibrational entropy, their free energy is different.  Open lattices, which exhibit lower free energy, can be selected in the system of patchy particles due to these entropic effects.

%%%%%%%%%%%%%%%%%%%%%%%%%%%%%%%%%%%%%%
\subsubsection{Effective Hamiltonian, mechanical stability, and harmonic approximation}
The rotational entropy discussed above can be plugged into the equation for the effective Hamiltonian~\eqref{EQ:HEF}, which is a version of the Ginzburg-Landau free energy with orientational degrees of freedom integrated out but positional degrees of freedom of the particles kept.  The rotational entropy is related to the orientational microscopic states we calculated above through
\begin{align}
	s_j = k_B \ln \MS_j ,
\end{align}
where $\MS_j $ is the microscopic states calculated for the $j$-th particle following the methods we discussed.

The resulting $\HEF$ is an effective Hamiltonian that only depends on particle positions.  In particular, with these rotational entropy terms, more constraints (in the Language of Maxwell's counting) are being introduced into the system comparing to the original Hamiltonian which has potential energy only.  For example, in the case of kagome lattices assembled from triblock Janus particles, the rotational entropy terms introduce constraints on the bond angles, which were the floppy modes in the potential-energy only Hamiltonian, as we discussed before.  Thus as a result the rotational entropy may in general provide mechanical stability to the open lattices, which are not stable at $T=0$ from potential energy considerations.  

To obtain the free energy for each lattice structure, we need to also integrate out the positional fluctuations.  In the mean time, it is also important to characterize the vibrational modes in these assembled lattices, which provide a valuable information in studying the mechanical response of the assembled structure.  In order to achieve this, we choose to use the harmonic approximation, i.e., keeping the effective Hamiltonian to quadratic order in displacement vectors $\vec{u}_j$ of particles, which facilitate analytic calculations and is convenient in extracting asymptotic behaviors.  A full calculation of the free energy involves more realistic modeling of the interaction potentials, including the hydrophobic and screened electrostatic interactions between the particles, which are anharmonic.  We shall investigate this full calculation numerically in our future study.

In order to obtain a harmonic form of the effective Hamiltonian, we shall make approximations on the rotational entropy so that it can be written into a form involving only quadratic terms in $\vec{u}_j$.  Given the special nature of the patchy particles, as shown in Fig.~\ref{FIG:RotEnt}, the rotational entropy is linear at small $\Delta\alpha$ but nonlinearity rises at slightly larger $\Delta\alpha$.  In practice the fluctuation of the bond-angle $\alpha$ can be relatively large owning to the low stiffness of the corresponding vibrational mode, as we shall discuss later.  Thus, it is a good approximation to require that the second moment $\langle (\Delta\alpha)^2 \rangle$ of the bond-angle fluctuations controlled by the rotational entropy in the original form should be kept the same in the harmonic approximation.  For the case of the elongated patches we have
\begin{align}
	\langle (\Delta\alpha)^2 \rangle 
	&=\frac{\int (\Delta\alpha)^2 e^{s/k_B} \, d\Delta \alpha }{ \int  e^{s/k_B} \, d\Delta \alpha }\nonumber\\
	&= \frac{2}{3} \left( \phi - \frac{\pi}{6}\right)^2 ,
\end{align}
where $s$ is given by Eq.~\eqref{EQ:sElon}.  This average can either be viewed as weighted by the effective Gibbs factor from the entropic term of the rotation of the given particle in the effective Hamiltonian~\eqref{EQ:HEF}, or as weighted by the number of microscopic states of the particle, because $e^{s/k_B}=\MS$.  The corresponding harmonic form with the same second moment is then
\begin{align}\label{EQ:SELH}
	-Ts &\simeq -\frac{\kappa}{2} (\Delta \alpha)^2 , \nonumber\\
	\kappa &= \frac{3k_B T}{2\left( \phi - \frac{\pi}{6}\right)^2} ,
\end{align}
where the resulting bond-angle stiffness $\kappa$ is proportional to $k_B T$ because it is of entropic origin.  
For the case of circular patches we can do a similar calculation, the result will be a modified value of the bond-angle rigidity $\kappa$.  The case discussed in Sec.~\ref{SEC:GAMMA} is different, because of the additional dependence on the bond angles $\g_1$ and $\g_2$.  We discuss the harmonic approximation of the rotational entropy of this case in App.~\ref{APP:Harmonic}, and the result is
\begin{align}\label{EQ:SLong}
	-Ts \simeq -\frac{\kappa}{2} \left(\Delta \alpha +\frac{\Delta\g_1 + \Delta\g_2}{2}\right)^2 ,
\end{align}
where $\kappa$ is the same as given in Eq.~\eqref{EQ:SELH}.  

%%%%%%%%%%%%%%%%%%%%%%%%%%%%%%%%%%%%%%
\subsubsection{Dynamical matrix of the lattices}
With the harmonic forms of the rotational entropy derived above, in order to write the whole the effective Hamiltonian~\eqref{EQ:HEF} into a harmonic form and thus express it in terms of a dynamical matrix, we also need to characterize the central force interactions $\HCF$ and make harmonic approximations on this part too.

The central force interactions between particles facing each other in their attractive patches are short-range attractions.  In the experiment of Refs.~\cite{Chen2011,Mao2013} they are of hydrophobic nature.  In order to make a harmonic approximation we expand this short-range attraction near its minimum to quadratic order, which is equivalent to a harmonic spring, and the potential can be written as
\begin{align}\label{EQ:APote}
	V_{i,j} = \frac{k}{2} \left(\vert \vec{R}_j - \vec{R}_i  \vert - l \right)^2 ,
\end{align}
where we assume the particles undergo a displacement from original positions $\vec{r}_i$ to displaced positions $\vec{R}_i$
\begin{align}
	\vec{r}_i \to \vec{R}_{i}=\vec{r}_i+\vec{u}_i ,
\end{align}
and $k$ is the spring constant, $l$ is the rest length.

For particles not facing each other both in their attractive patch the central force interaction is essentially a hard sphere repulsion.  In practice the repulsion is of a very small range.  In the experiment of Refs.~\cite{Chen2011,Mao2013} this repulsion is of the nature of strongly-screened electrostatic repulsion with the range of repulsion comparable to surface roughness.  We can also expand this interaction around the stable inter-particle distances in the lattice to quadratic order and approximate it as a harmonic spring.  Because this interaction is purely repulsive in nature, the expansion in terms of distance variations exhibit a linear term, which contribute to the internal stress of the lattice but do not change our analysis of vibrational modes.  Similar to the attractive bonds, the potential of these non-attractive bonds can be written as
\begin{align}\label{EQ:RPote}
	V_{i,j} = \frac{k_r}{2} \left(\vert \vec{R}_j - \vec{R}_i  \vert - l \right)^2 ,
\end{align}
where $k_r$ is the spring constant for the non-attractive bonds, $l$ is the rest length.

We can then expand the central-force potentials in Eqs.(\ref{EQ:APote},\ref{EQ:RPote}) in the displacement vectors $\vec{u}_i$ and keep to quadratic order.  The same procedure needs to be applied to the rotational entropy terms through the relation of bond angle with displacement vectors in $2D$
\begin{align}
	\Delta \alpha_{i} =\frac{1}{a}(\hat{r}_{hi}\times \vec{u}_{hi} - \hat{r}_{ij}\times \vec{u}_{ij})
	\cdot \hat{z}
\end{align}
where $h,i,j$ labels the particles forming the bond angle, $\hat{r}_{hi}$ and $\hat{r}_{hi}$ are the unit vectors pointing from $h$ to $i$, and from $i$ to $j$ in the undeformed state.  The unit vector $\hat{z}$ points in the third dimension out of the plane.  The diameter of the particle, which is the same as the distance between nearest neighbors, is denoted by $a$.  The same formula applies to $\Delta\g_1, \Delta\g_2$ too.

Thus the effective Hamiltonian, which contains both the central-force terms and the rotational-entropy terms
\begin{align}\label{EQ:HEFFull}
	\HEF =& \sum_{\langle i j\rangle, a} \frac{k}{2} \left(\vert \vec{R}_j - \vec{R}_i  \vert - l \right)^2 \nonumber\\
	  &+\sum_{\langle i j\rangle, r} \frac{k_r}{2} \left(\vert \vec{R}_j - \vec{R}_i  \vert - l \right)^2 \nonumber\\
	  &+ \frac{\kappa}{2} \sum_{i} \left(\Delta \alpha_i + g\frac{\Delta\g_{1,i} + \Delta\g_{2,i}}{2}\right)^2 ,
\end{align}
where the summation $\langle i j\rangle, a$ is over attractive bonds, $\langle i j\rangle, r$ is over non-attractive bonds, $i$ in the third term is over all particles.  The parameter $g=0$ if we adopt the simple form~\eqref{EQ:SELH} of the rotational entropy and $g=1$ if we adopt the form~\eqref{EQ:SLong} which allows bond-length fluctuations.  In calculating the figures in this Section and the next, we used $g=1$.  The result will be qualitatively the same if we use $g=0$.  This effective Hamiltonian of the lattice can be readily written in the formalism of dynamical matrix as
\begin{align}\label{EQ:HEFD}
	\HEF = \frac{1}{2} \sum_{\ell,\ell'} \vec{U}_{\ell} \mathbf{D}_{\ell,\ell'} \vec{U}_{\ell'} ,
\end{align}
where $\ell$ labels unit cells each containing $m$ particles, and $\vec{U}_{\ell}$ denote the $2m$-dimensional displacement vector of the particles in the unit cell in two dimensions.  In App.~\ref{APP:DM} we derive the dynamical matrix for the hexagonal and the kagome lattices.

\subsubsection{Normal modes of the lattices}
Normal modes of the two lattices can be directly derived from their dynamical matrices $\mathbf{D}_{\ell,\ell'}$ as eigenmodes of the matrices.  For periodic lattices this analysis is most conveniently done in momentum space using $\mathbf{D}_{\vec{q},\vec{q}'}$.
\begin{figure}
	\centering
		\subfigure[]{\includegraphics[width=.45\textwidth]{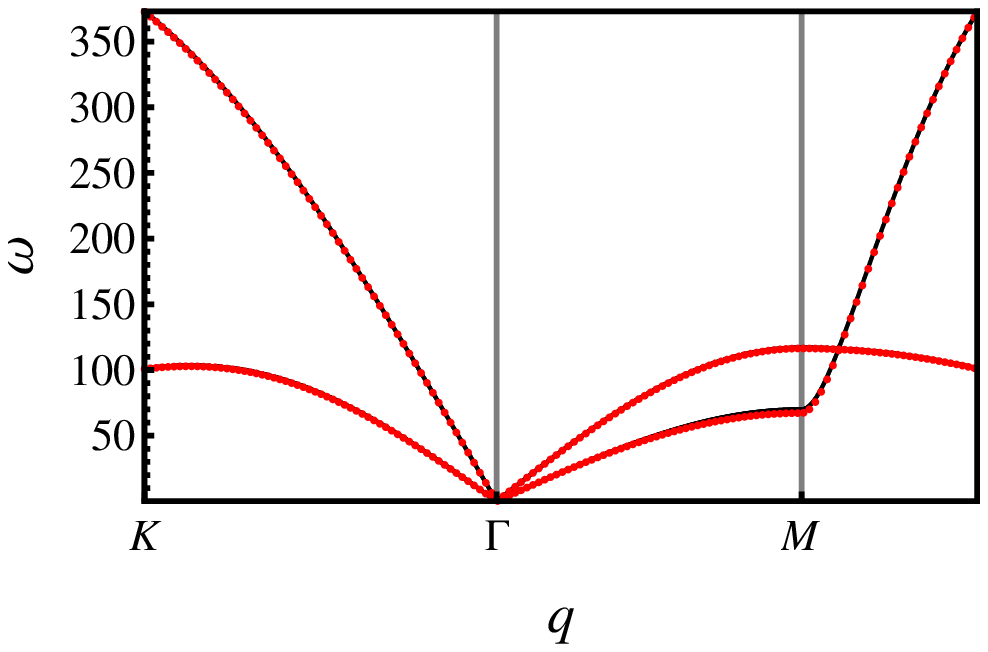}}\\
		\subfigure[]{\includegraphics[width=.45\textwidth]{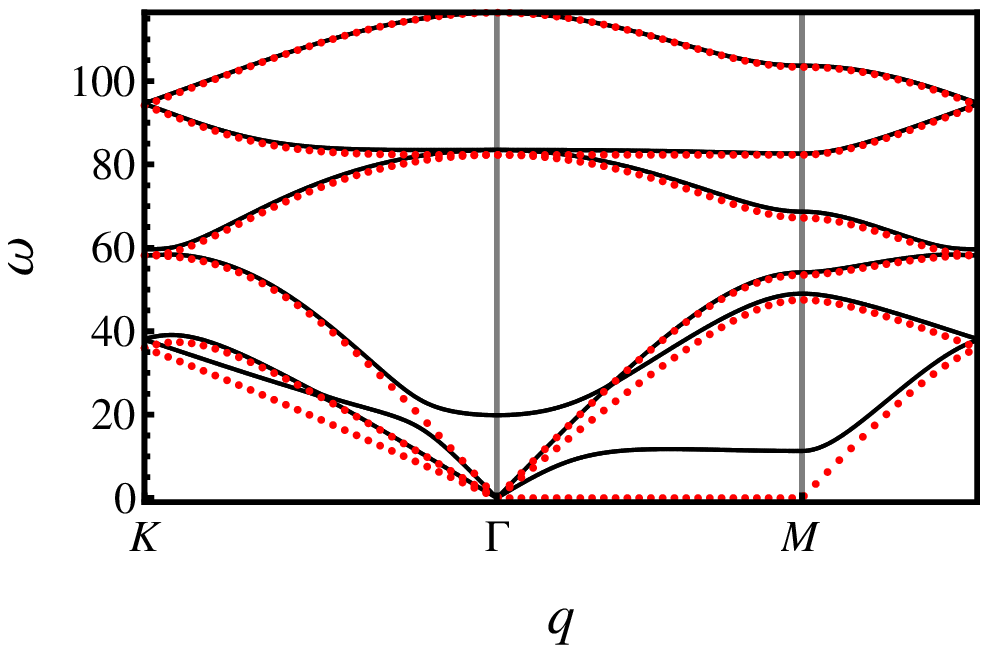}}\\
		\subfigure[]{\includegraphics[width=.12\textwidth]{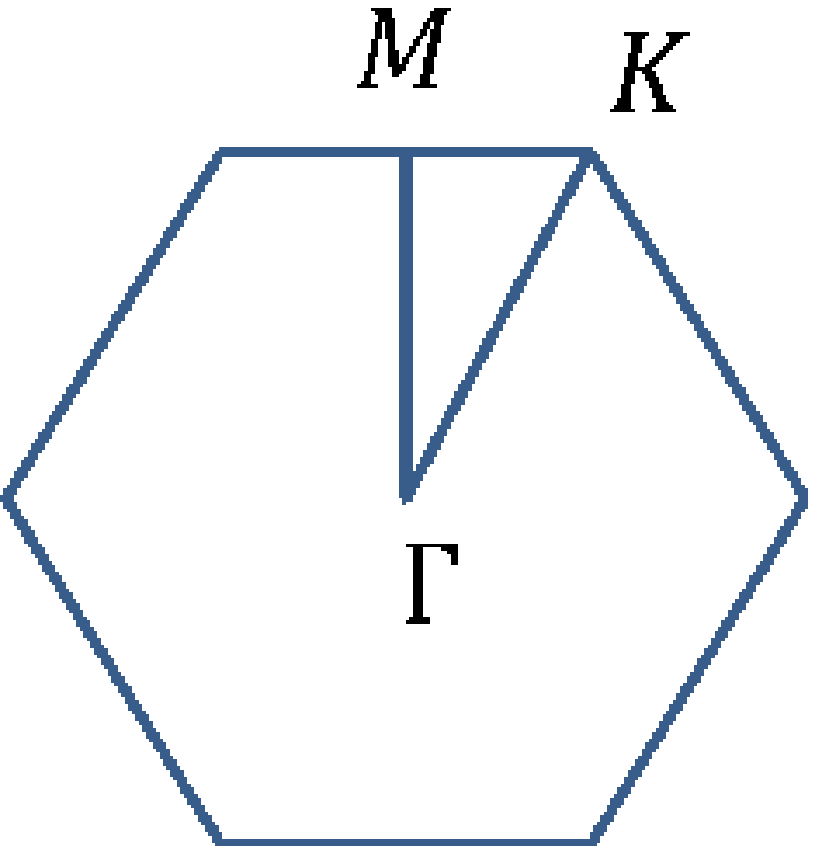}}
	\caption{Phonon dispersion relations for the hexagonal lattice (a) and the kagome lattice (b).  The dispersion relations are plotted in the first Brillouin zone.  The black solid curves represent the case with $\kappa=0$, $k=2300k_B T a^{-2}$ and $k_r=20k$.  The red dotted curves represent the case with $\kappa=33k_B T$, $k=2300k_B T a^{-2}$ and $k_r=20k$.  The hexagonal lattice show relatively high frequency in some part of the dispersion relation plot because of the high value of $k_r$.  The first Brillouin zone for the hexagonal and the kagome lattices are shown in (c).}
	\label{FIG:SPEC}
\end{figure}
 
For the hexagonal lattice, because $\mathbf{D}_{\vec{q},\vec{q}'}$ is $2\times 2$, there are two branches of modes as shown in Fig.~\ref{FIG:SPEC}a, corresponding to the two acoustic phonons in two dimensions.  The frequencies of both of the two branches are controlled by $k$, $k_r$, and $\kappa$.  In this plot we used the values of $k$, $k_r$, and $\kappa$ from fitting to the experiment as discussed in Ref.~\cite{Mao2013} and also the case with $\kappa=0$.  It is straightforward to see that the difference in the frequency in the case with $\kappa=0$ and $\kappa>0$ (but small compare to $ka^2$) is small and only quantitative in the hexagonal lattice.

For the kagome lattice, there are six branches of modes as shown in Fig.~\ref{FIG:SPEC}b, because each unit cell consists three particles.  As discussed in Ref.~\cite{Mao2011a}, there are two acoustic branches and four optical branches of phonons.  In the case of $\kappa=0$ the lowest optical branch drops to zero frequency (upon hybridization with transverse phonons) along the $\Gamma M$ direction in the first Brillouin zone.  This branch corresponds to the floppy modes of the lattice, which are of zero-frequency if there is only central force interactions between nearest neighbors, as shown in Fig.~\ref{FIG:FM}.  For $\kappa>0$ this branch gains a gap that is proportional to $\sqrt{\kappa}$.  Thus the two cases of $\kappa=0$ and $\kappa>0$ are qualitatively different for the kagome lattice.  In particular, because of the $\ln \omega$ contribution to the free energy, as $\kappa$ becomes small, the free energy of the kagome lattics is significantly lowered, as we will discuss in Sec.~\ref{SEC:Results}.

In addition, the eigenvalues (square of the frequencies) of the six modes of the kagome lattice at $\vq=0$ are given by 
\begin{align}\label{EQ:zeroF}
	\{ 0,0,\frac{12\kappa}{a^2},3k+\frac{3(1+3g^2)\kappa}{2a^2},3k+\frac{3(1+3g^2)\kappa}{2a^2},6k \} 
\end{align}
where the floppy mode is the third one in the list.

%%%%%%%%%%%%%%%%%%%%%%%%%%%%%%%%%%%%%%
%%%%%%%%%%%%%%%%%%%%%%%%%%%%%%%%%%%%%%
%%%%%%%%%%%%%%%%%%%%%%%%%%%%%%%%%%%%%%
\section{Results: Competition between open and close packed colloidal lattices}
\label{SEC:Results}
%%%%%%%%%%%%%%%%%%%%%%%%%%%%%%%%%%%%%%
%%%%%%%%%%%%%%%%%%%%%%%%%%%%%%%%%%%%%%
\subsection{Free energy difference}
At finite temperature and in thermal equilibrium, the relative stability of different lattice phases is determined by their free energies per particle $f$.  The free energy per particle of a lattice can be directly evaluated using the partition function~\eqref{EQ:ZEF} and the effective Hamiltonian~\eqref{EQ:HEFD} we developed above,
\begin{align}
	f&=-(Nm)^{-1} k_B T \ln Z \nonumber\\
	&= \frac{k_B T}{2Nm} \ln \,\textrm{Det} \mathbf{D} -\frac{d}{2}k_B T \ln \,\left(\frac{2\pi k_B T}{a^2}\right)
\end{align}
where the second term is a constant which we shall denote as $\tilde{C}$.  The determinant is taken over the $mNd$ dimensions of the tensor $D_{\ell,\ell}^{\alpha\beta}$, where $\ell,\ell'$ label the unit cells and $\alpha\beta$ label the Cartesian indices.

We can further write this free energy using the dynamical matrix in momentum space as defined in Eqs.~(\ref{EQ:Dqq},\ref{EQ:DqqtoDq},\ref{EQ:Dq}), in which it is diagonal in $\vq$ and convenient to evaluate
\begin{align}
	f&=\frac{k_B T}{2Nm} \ln \,\textrm{Det} \mathbf{D} +\tilde{C}\nonumber\\
	&= \frac{k_B T}{2Nm} \ln \,\Big\lbrack  (V v_0)^{-Nm}
		\textrm{Det} \tilde{\mathbf{D}} \Big\rbrack +\tilde{C}\nonumber\\
	&= \frac{k_B T}{2Nm} \ln \,\textrm{Det} \mathfrak{D} +\tilde{C}\nonumber\\
	&= \frac{k_B T}{2m} \frac{v_0}{V} \sum_{\vq} \ln \det \mathfrak{D}_{\vq} +\tilde{C},
\end{align}
where the determinant in the last line is only taken in the $2m$ dimensional space at a given $\vq$.  In the continuum limit we can write this into
\begin{align}\label{EQ:fint}
	f=\frac{k_B T}{2m} \int_{1BZ}\frac{d^{d}\vq}{(2\pi)^d v_0^{-1}}  
	\ln \det \mathfrak{D}_{\vq} +\tilde{C}
\end{align}
where the integral is over the first Brillouin zone.  In what follows we specialize to the free energies of kagome and hexagonal lattices assembled from triblock Janus particles.

Firstly, we consider the kagome lattice.  In the $\kappa=0$ case, the floppy modes along $\Gamma M$ lines lead to a logarithmic divergence in the $\ln \det \mathfrak{D}_{\vq}$ terms in the integrand.  However the free energy which is after the integral over the first Brillouin Zone does converge to a finite value,
\begin{align}
	f_{K} \vert_{\kappa=0}
	= k_B T \left\lbrack \frac{d}{2}\ln k + \frac{1}{2}\ln\left(\frac{3}{4}\right) \right\rbrack +\tilde{C},
\end{align}
where $f_K$ denotes the free energy per particle of the kagome lattice.  It is noteworthy that the dependence on $k$ is through an additive term, because in the case of $\kappa=0$ all terms in $\mathfrak{D}$ is proportional to $k$ and thus $k$ can thus be factorized out.  This term can be combined with $\tilde{C}$ leading to a term baring the dimension of energy
\begin{align}
	C=\frac{d}{2}k_B T \ln \,\left(\frac{ka^2}{2\pi k_B T} \right) ,
\end{align}
so that
\begin{align}
	f_{K} \vert_{\kappa=0}
	= \frac{k_B T}{2} \ln\left(\frac{3}{4}\right)  +C .
\end{align}
For the comparison between different lattices consisting the same type of patchy particles, this term is a lattice-independent constant and will cancel out in the free energy differences.  

\begin{figure}
	\centering
		\includegraphics[width=.4\textwidth]{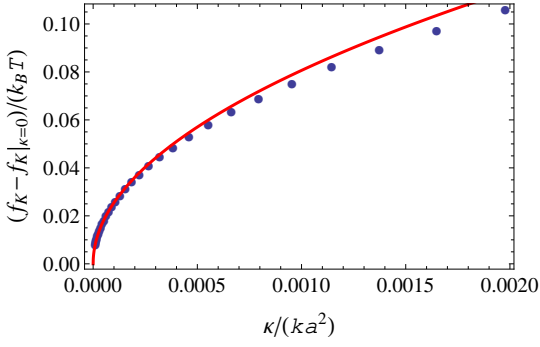}
	\caption{The increase of kagome lattice free energy per particle as a function of the bending stiffness $\kappa$.  Blue dots represent numerical integrals and the red line represent fitted form $\left(f_{K}-f_{K}\vert_{\kappa=0}\right)/(k_B T)\simeq b_0 \sqrt{\kappa/(ka^2)}$ with $b_0=2.55$.}
	\label{FIG:fKcorrection}
\end{figure}

For the case of $\kappa>0$ but small, the leading order correction to $f_{K}$ is proportional to $\sqrt{\kappa/(ka^2)}$, as shown in Fig.~\ref{FIG:fKcorrection}.  This non-analytical dependence on $\kappa$ comes from the logarithmic divergence in $\ln \omega$, which originates from the mechanical instability of the kagome lattice.

\begin{figure}
	\centering
		\subfigure[]{\includegraphics[width=.4\textwidth]{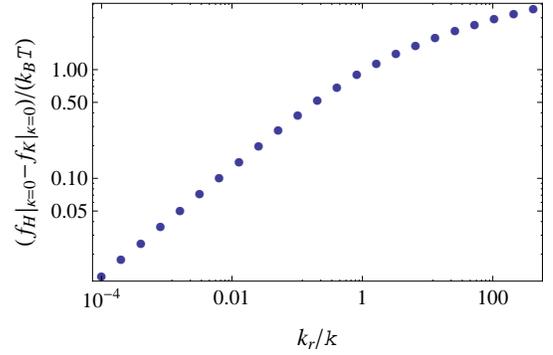}}
		\subfigure[]{\includegraphics[width=.4\textwidth]{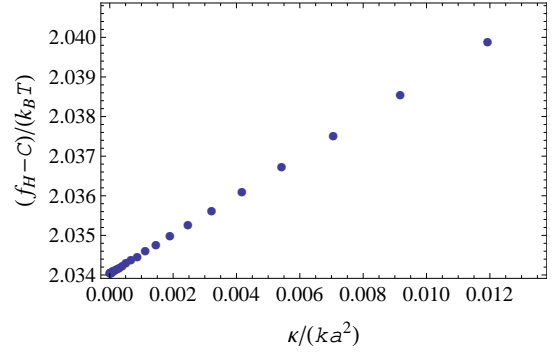}}
	\caption{(a) The difference between the hexagonal lattice and the kagome lattice free energy per particle as a function of $k_r/k$ at $\kappa=0$.  At $k_r/k=0$ and $\kappa=0$ there is no difference between these two lattices so the curve goes to zero.  For $k_r/k$ of order $1$ or greater, the difference between the two free energies is of the order of $1k_B T$.  (b) The hexagonal lattice free energy per particle as a function of $\kappa/(ka^2)$ at $k_r/k=20$.  It is clear that in this case the correction due to $\kappa$ is linear.}
	\label{FIG:fH}
\end{figure}

Secondly, we consider the free energy of the hexagonal lattice.  In the case of $\kappa=0$, the lattice is stable and there is no divergence in the integrand at all.  Owing to the extra parameter $k_r$ associated with the non-attractive bonds, the free energy per particle of the hexagonal lattice $f_{H} \vert_{\kappa=0}$ depends on both $k$ and $k_r$.  In particular, given the logarithm, $k$ can be factorized out contributing to the same constant $C$ as in the kagome lattice.  The other term, instead of being a constant, bares a dependence on the ratio $k_r/k$.  In the limit of $k_r=0$ this just becomes the same problem as the kagome lattice with $\kappa=0$.  In real systems typically $k_r$ is of the same order of magnitude as or greater strength than $k$.  In the experiment in Ref.~\cite{Mao2013} it is found that $k_r/k\simeq 20$.  The dependence of $f_{H} \vert_{\kappa=0}$ on $k_r/k$ is shown in Fig.~\ref{FIG:fH}(a).  

For the case of $\kappa>0$, $f_{H}$ increases with $\kappa$ with a leading order correction proportional to $\kappa/(ka^2)$ instead of $\sqrt{\kappa/(ka^2)}$ (except for the case in which $k_r$ is very small), because the hexagonal lattice is not affected by the marginal stability as is the kagome lattice.  This is shown in Fig.~\ref{FIG:fH}(b).

\begin{figure}
	\centering
		\includegraphics[width=.4\textwidth]{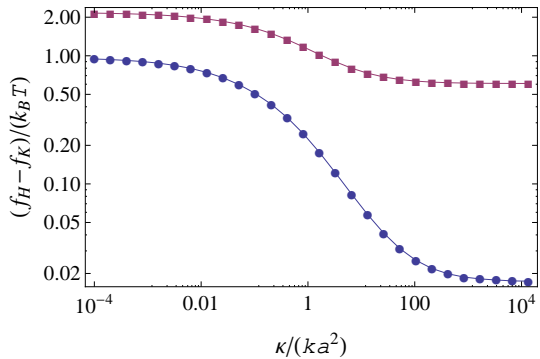}
	\caption{Free energy difference between the hexagonal and the kagome lattices as a function of $\kappa/(ka^2)$ at $k_r=1$ (lower blue curve) and $k_r=20$ (upper red curve).}
	\label{FIG:Deltaf}
\end{figure}

Therefore, in calculating the difference between the free energy per particle of the hexagonal and the kagome lattices, the constant term $C$ cancels out, and the resulting difference depends on the two ratios $\kappa/(ka^2)$ and $k_r/k$.  In particular, $\Delta f=f_H - f_K$ decreases with $\kappa/(ka^2)$, and increases with $k_r/k$, as shown in Fig.~\ref{FIG:Deltaf}.

From this calculation we see that the free energy difference is significant, i.e., of order $k_B T$, in two situations: (i) $\kappa/(ka^2)\ll 1$, or (ii) $k_r/k \gg 1$.  This provide us with the criterion of the selection of the kagome lattice over the hexagonal lattice by free energy considerations.  For the case of $k_r/k$ not significantly greater than unity, the selection of the kagome lattice is only effective when $\kappa/(ka^2)\ll 1$.  Because the bending stiffness $\kappa$ originates from the rotational entropy as shown in Eq.\eqref{EQ:SELH}, this condition leads to a characteristic patch size
\begin{align}
	\hps_c = \frac{\pi}{6} + \sqrt{\frac{3k_B T}{2k a^2}}
\end{align}
above which the selection of the kagome lattice is effective (if $k_r/k$ not already significantly greater than unity).  This reveals the perhaps surprising effect that the selection of open lattices can be enhanced by having excessively large attractive patch~\cite{Mao2013}.

%%%%%%%%%%%%%%%%%%%%%%%%%%%%%%%%%%%%%%
%%%%%%%%%%%%%%%%%%%%%%%%%%%%%%%%%%%%%%
\subsection{Phase diagram}
At positive pressure there exist a first-order transition from the kagome to the hexagonal lattice, because the hexagonal lattice is of higher packing fraction~\cite{Mao2013}.  This transition can be described by transform from the statistical ensemble of fixed lattice structure into the ensemble of fixed pressure through a Legendre transformation
\begin{align}
	g =f+p v
\end{align}
where $g$ is the Gibbs free energy per particle, $p$ is the pressure, and $v=V/N$ is the mean area occupied by one particle in the given lattice.  At fixed $p$ in equilibrium, the phase with lower $g$ is more stable.  Therefore we can determine the  boundary between the kagome and the hexagonal phases using the equal $g$ line.

For a qualitative description we make further simplification that within a phase the change in the lattice constant as a function of pressure can be ignored, thus the only volume change comes from the difference between the two lattice structures.  In particular the value of $v$ in the kagome and the hexagonal lattices are respectively
\begin{align}
	v_{K} = \frac{2a^2}{\sqrt{3}}, \quad v_{H} =  \frac{\sqrt{3}a^2}{2} ,
\end{align}
and we define $\Delta v=v_H - v_K$.

We also make the simplifying assumption that the stiffness parameters $k$, $k_r$ and $\kappa$ also remain the same as pressure changes, because the lattice constant is unchanged.  Thus the difference in Helmholtz free energy between the two lattices, $\Delta f=f_H - f_K$, is the same at different pressures.  In addition, the potential energy is the same in the two phases, because each particle have $4$ attractive bonds in both phases, and as a result, we have $\Delta f = -T \Delta s $ where $\Delta s$ is the entropy difference per particle between the two lattices.

\begin{figure}
	\centering
		\includegraphics[width=.4\textwidth]{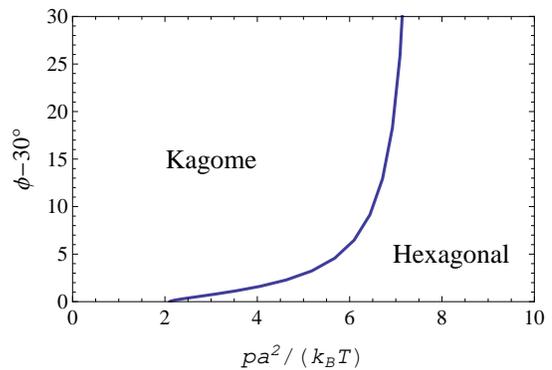}
	\caption{Equilibrium phase diagram showing the boundary between the hexagonal and the kagome lattice determined by the equal Gibbs free energy per particle line.}
	\label{FIG:PD}
\end{figure}

Therefore the equal $g$ line, which describes the phase boundary, correspond to the equation
\begin{align}
	p = T \frac{\Delta s}{\Delta v} .
\end{align}
This leads to the phase diagram shown in Fig.~\ref{FIG:PD}.  If this phase diagram is plotted in the traditional $p-T$ plane, the phase boundary will be a straight line, of which the slope positively depend on the patch-size.  In addition, at high temperature we expect that both lattices will melt into a fluid phase, which is not considered in this theory.

%%%%%%%%%%%%%%%%%%%%%%%%%%%%%%%%%%%%%%
%%%%%%%%%%%%%%%%%%%%%%%%%%%%%%%%%%%%%%
%%%%%%%%%%%%%%%%%%%%%%%%%%%%%%%%%%%%%%
\section{Discussion}\label{SEC:Discussion}
In this Paper we construct a generic theory based on equilibrium statistical mechanics and lattice dynamics for the self-assembly of periodic lattices from patchy particles.  We discuss the entropic effects in open lattices formed by patchy particles.  In particular, we show that the rotational entropy can provide mechanical stability to open lattices and vibrational entropy can lower the free energy of the open lattice relative to that of close-packed lattices.  These effects are essential to the stabilization and selection of open lattices which are not stable in pure potential energy considerations.

The predictions of our theory agrees well with experiments on the triblock Janus particle system~\cite{Mao2013}.  In particular, the transition between the hexagonal and the kagome lattices predicted by this theory has been confirmed in experiment with increased lateral pressure.  The generic dependence of the phase boundary between the open and close-packed lattices on temperature and patch-size has yet to be confirmed experimentally.

This stabilization and selection effect of entropy in patchy particle systems is an example of the \lq\lq order by disorder\rq\rq  effect, in which degenerate states at $T=0$ gain different free energy due to thermal or quantum fluctuations~\cite{Villain1980,Henley1987,Chubukov1992,Reimers1993,CastroNeto2006}.  Related examples of this effect has been discussed in homogeneous colloids.  For example, close-packed lattice structures can be favored in hard-sphere homogeneous-colloid systems by entropy, because the regular lattice structure allows more \lq\lq rattle room\rq\rq  for the particles~\cite{Pusey1986}.  For patchy colloids, as discussed here, the entropic effect is even more interesting owing to the nontrivial rotational degrees of freedom.

In order to arrive at the form of the effective Hamiltonian that is suitable for harmonic lattice dynamics, a series of approximations have been made in this theory, and the major ones are: (i) the pair-wise interactions is very short-ranged and is of step-function nature, i.e., flat in attractive and non-attractive surfaces areas and only changes abruptly as particle orientation is altered across the boundary; (ii) both the central-force inter-particle potential and the effective bending potential term from rotational entropy can be approximated as harmonic in displacement vectors $\vec{u}$ of the particles.

The approximation (i) is made so that the orientational degrees of freedom of different particles can be decoupled.  For the experiment in Refs.~\cite{Chen2011,Mao2013} the salt concentration is high enough to screen the length-scale of electrostatic repulsion down to the scale of surface roughness and the approximation (i) is valid in these cases.  For experiments in which the pair-wise potential is not so flat as a function of particle orientations, the orientational degrees of freedom of different particles can not be decoupled.  Furthermore, the kagome lattice may directly gain some stability from potential energy.  Nevertheless, typically for patchy particles this potential energy variation as a function of bond angles is not strong enough to actually stabilize the lattice, and the effect of rotational entropy on the stability may still dominate.

The approximation (ii) validates the application of lattice dynamics to the effective Hamiltonian of the system after orientational degrees of freedom are integrated out.  We made this approximation to obtain qualitative predictions of the free energies of different lattices, which reveal the generic feature of the phase diagram without requiring complicated calculations.  More accurate calculation of the free energies of the lattices can be done numerically with more realistic modeling of the inter-particle potentials.  We do not expect qualitative difference in the conclusion by replacing the harmonic approximation with more realistic potentials.

This theory is readily generalizable to more cases of self-assembly of patchy particles, including three-dimensional lattices.  From the entropic stabilization mechanism discussed in this theory, it is expectable that in three dimension, transitions between close-packed face-centered-cubic and open pyrochlore/perovskite lattices occur in the tri-block patchy particle system, in a similar fashion as the hexagonal/kagome transition in two dimensions.  This entropic effect provides a mechanism in which regular periodic lattices are automatically favored with remarkably simple designs of building blocks.  

\begin{acknowledgments}
We are grateful for fruitful discussions with Qian Chen, Steve Granick, and Tom C. Lubensky.  
This work was supported in part by National Science Foundation under DMR-1104707.
\end{acknowledgments}

%%%%%%%%%%%%%%%%%%%%%%%%%%%%%%%%%%%%%%
%%%%%%%%%%%%%%%%%%%%%%%%%%%%%%%%%%%%%%
%%%%%%%%%%%%%%%%%%%%%%%%%%%%%%%%%%%%%%
\appendix

%%%%%%%%%%%%%%%%%%%%%%%%%%%%%%%%%%%%%%
\section{Harmonic approximation of the rotational entropy}\label{APP:Harmonic}
In this section we consider the Harmonic approximation of the rotational entropy
\begin{align}
	s=k_B \ln \MS,
\end{align}
with $\MS$ given in Eq.~\eqref{EQ:REC}.  Due to the dependence on the angles $\g_1, \g_2$ it is not possible to always keep the second moment $\langle (\Delta\alpha)^2 \rangle$ invariant in this approximation.  Therefore we need to generalize our method of approximation.  

Here we show that the harmonic approximation we defined for the simple case of $\MS$ defined in Eq.~\eqref{EQ:MSS} can also be obtained through the following more generic procedure.  We firstly define the Fourier transform of the statistical weight $e^{s/k_B}$
\begin{align}
	y(q)=\int e^{s/k_B} \, e^{-iq \Delta\alpha} d\Delta \alpha .
\end{align}
Then we can take the logarithm of $y(q)$ and keep to quadratic order in $q$.  This leads to a Gaussian approximation of $y(q)$ as
\begin{align}
	\tilde{y}(q) = y(0) e^{-\frac{q^2}{2\kappa}}.
\end{align}
The inverse Fourier transform of $\tilde{y}(q)$ leads us to a Gaussian approximation of the original statistical weight
\begin{align}
	e^{\tilde{s}/k_B} = \int \frac{dq}{2\pi} \,\tilde{y}(q) e^{iq\Delta\alpha} = e^{-\frac{\kappa \Delta \alpha^2}{2}} .
\end{align}
It is straightforward to show that following such a procedure of approximating a statistical distribution into a Gaussian distribution, the second moment $\langle (\Delta\alpha)^2 \rangle$ is kept invariant.  The second moment can be calculated using the full Fourier transform $y(q)$ as follows
\begin{align}
	\langle (\Delta\alpha)^2 \rangle&= \int d\Delta\alpha  (\Delta\alpha)^2 e^{s/k_B} 
	/ \int d\Delta\alpha  e^{s/k_B} \nonumber\\
	&= \frac{1}{y(0)} \int d\Delta\alpha  \int \frac{dq}{2\pi} \, y(q) e^{iq\Delta\alpha} \nonumber\\
	&= \frac{y''(0)}{y(0)} .
\end{align}
Thus it is clear for this second moment we only need to keep to $O(q^2)$ on the exponent of $y(q)$, which is $\tilde{y}(q)$, and higher order terms will not contribute.   

Now we can use this method to calculate the harmonic approximation of $\MS$ given in Eq.~\eqref{EQ:REC}, which applies to the case with bond-length fluctuations allowed.  We have the Gaussian form of the Fourier transform
\begin{align}
	\tilde{y}(q) = y(0) e^{-\frac{\beta^2 q^2 + 6i(\Delta\g_1+\Delta\g_2)q}{12}}
\end{align}
in which we also kept to quadratic order in $\Delta\g_1,\Delta\g_2$ because they are small fluctuations.  Transforming back we arrive at
\begin{align}
	-Ts \simeq -\frac{\kappa}{2} \left(\Delta \alpha +\frac{\Delta\g_1 + \Delta\g_2}{2}\right)^2 ,
\end{align}
where $\kappa$ is the same as given in Eq.~\eqref{EQ:SELH}.

%%%%%%%%%%%%%%%%%%%%%%%%%%%%%%%%%%%%%%%%%%%%%%
\section{Dynamical matrices and normal modes of the effective Hamiltonian}
\subsection{Constructing the dynamical matrices}\label{APP:DM}
\subsubsection{The hexagonal lattice}
\begin{figure}
	\centering
		\subfigure[]{\includegraphics[width=.32\textwidth]{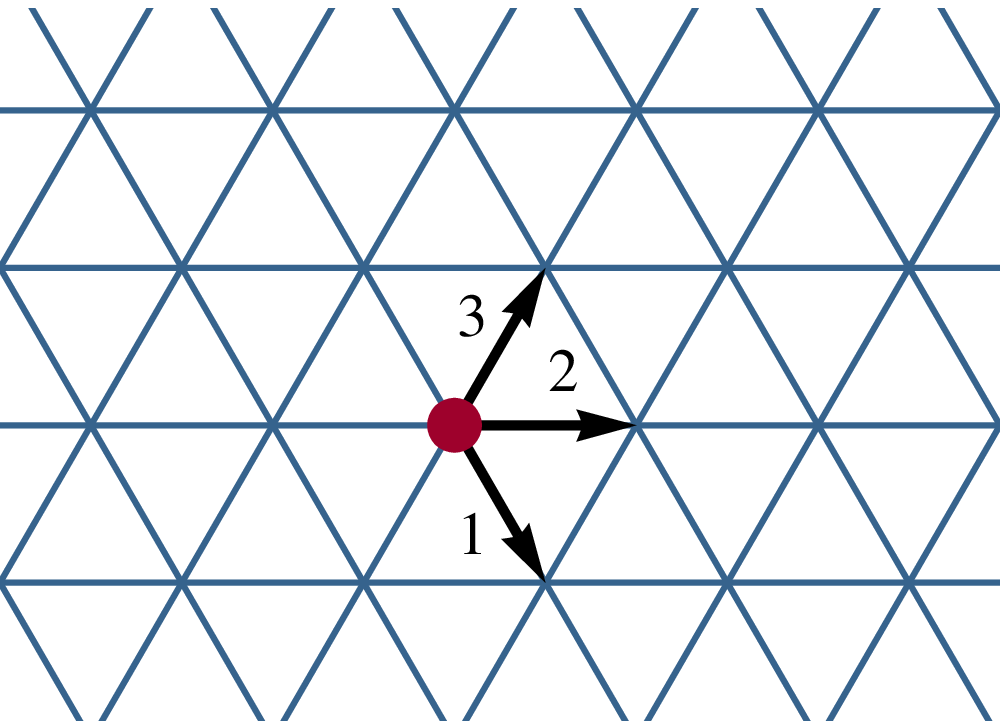}}
		\subfigure[]{\includegraphics[width=.32\textwidth]{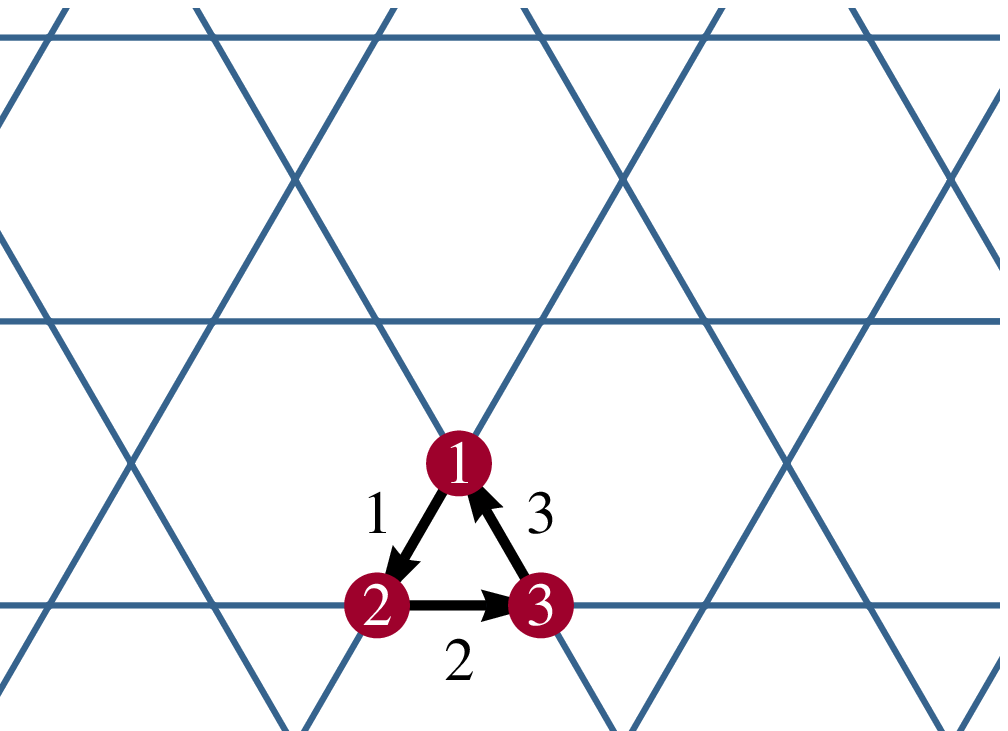}}
	\caption{The architecture and vectors in hexagonal (a) and kagome (b) lattices. The black numbers $1,2,3$ marks the three unit vectors $\uvepara_n$ in the two lattices.  The red disks with white numbers on them in (b) label the $3$ particles in a unit cell in the kagome lattice.}
	\label{FIG:Latt}
\end{figure}

Starting from the effective Hamiltonian as given in Eq.~\eqref{EQ:HEFFull}, we can construct the dynamical matrix of the hexagonal lattice following the method discussed in Ref.~\cite{Mao2011}, 
\begin{align}\label{EQ:hexagonalD}
	\mathfrak{D}_{\vec{q}}^{\textrm{hexagonal}} 
	=& k  \left(\vec{B}_{1,\vec{q}}^{s} 
	 \vec{B}_{1,-\vec{q}}^{s} +\vec{B}_{3,\vec{q}}^{s} \vec{B}_{3,-\vec{q}}^{s} \right) 
	 + k_r  \vec{B}_{2,\vec{q}}^{s}\vec{B}_{2,-\vec{q}}^{s} \nonumber\\
	 &+ \frac{\kappa}{a^2}  \left( \vec{B}_{\vec{q}}^{b} 
	 +g\frac{ \vec{\Gamma}_{1,\vec{q}}^{b}+ \vec{\Gamma}_{2,\vec{q}}^{b}}{2} \right) \nonumber\\
	 &\quad * \left( \vec{B}_{-\vec{q}}^{b} 
	 +g\frac{ \vec{\Gamma}_{1,-\vec{q}}^{b}+ \vec{\Gamma}_{2,-\vec{q}}^{b}}{2} \right)
\end{align}
where 
\begin{align}
	\vec{B}_{n,\vec{q}}^{s} &\equiv (1-e^{-i \vq \cdot \uvepara_n }) \uvepara_n \nonumber\\
	\vec{B}_{\vec{q}}^{b} &\equiv (1-e^{-i \vq \cdot \uvepara_1 }) \uveperp_1
	+(e^{-i \vq \cdot \uvepara_3 }-1) \uveperp_3 \nonumber\\
	\vec{\Gamma}_{1,\vec{q}}^{b} & \equiv -(1-e^{-i \vq \cdot \uvepara_1 }) \uveperp_1
	+(e^{-i \vq \cdot \uvepara_3 }-1) \uveperp_3 \nonumber\\
	\vec{\Gamma}_{2,\vec{q}}^{b} & \equiv (1-e^{-i \vq \cdot \uvepara_1 }) \uveperp_1
	-(e^{-i \vq \cdot \uvepara_3 }-1) \uveperp_3 
\end{align}
and 
\begin{align}
	\uvepara_n &\equiv \cos \lbrack(n-2)\pi/3\rbrack \hat{e}_x 
	+ \sin \lbrack(n-2)\pi/3\rbrack \hat{e}_y \nonumber\\
	\uveperp_n &\equiv -\sin \lbrack(n-2)\pi/3\rbrack \hat{e}_x 
	+ \cos \lbrack(n-2)\pi/3\rbrack \hat{e}_y \nonumber\\
\end{align}
are the unit vectors along or perpendicular to the bonds as shown in Fig.~\ref{FIG:Latt}a.  The lattice constant of this hexagonal lattice is equal to the bond length $a$.  The momentum space dynamical matrix is related to the real space one through
\begin{align}\label{EQ:Dqq}
	\tilde{\mathbf{D}}_{\vec{q},\vec{q}'} &=v_{0,\textrm{h}}^2 \sum_{\ell,\ell'}
	e^{-i\vq \cdot \vec{r}_{\ell}+i\vq' \cdot \vec{r}_{\ell'}} \mathbf{D}_{\ell,\ell'} , \nonumber\\
	 \mathbf{D}_{\ell,\ell'} &= \frac{1}{V^2} \sum_{\vec{q},\vec{q}'}
	e^{-i\vq \cdot \vec{r}_{\ell}+i\vq' \cdot \vec{r}_{\ell'}} \tilde{\mathbf{D}}_{\vec{q},\vec{q}'}
\end{align}
following the convention of Fourier transform on lattices in Ref.~\cite{Lubensky2000}.  The factor $v_{0,\textrm{h}}=\sqrt{3}a^2/2$ is the area of the unit cell for the hexagonal lattice.  Using translational invariance the dynamical matrix in momentum space can be written into
\begin{align}\label{EQ:DqqtoDq}
	\tilde{\mathbf{D}}_{\vec{q},\vec{q}'} = V v_{0,h}\, \delta_{\vec{q}-\vec{q}'} 
	\mathfrak{D}_{\vec{q}}
\end{align}
with
\begin{align}\label{EQ:Dq}
	\mathfrak{D}_{\vec{q}} = \sum_{\ell-\ell'}
	e^{-i\vq \cdot (\vec{r}_{\ell}- \vec{r}_{\ell'})} \mathbf{D}_{\ell,\ell'}
\end{align}

The bending part of this dynamical matrix is different from the version used in Ref.~\cite{Mao2011,Broedersz2011} for the reason that here each particle just has one bending energy term, and the two bonds forming this angle is $120^{\circ}$ in the reference state rather than $180^{\circ}$ as for the filaments.

\subsubsection{The kagome lattice}
The dynamical matrix for the kagome lattice can be constructed in a similar way.  Because there are $3$ particles in one unit cell of the kagome lattice, the displacement vector $\vec{U}_{\ell}$ is $6$ dimensional representing the $2d$ motion of the $3$ particles,
\begin{align}
	\vec{U}_{\ell} = \{u_{\ell,1,x},u_{\ell,1,y},u_{\ell,2,x},u_{\ell,2,y},u_{\ell,3,x},u_{\ell,3,y}\} .
\end{align}
To denote the directions of the bonds in the lattice, we define the unit vectors
\begin{align}
	\uvepara_n &\equiv \cos \lbrack(n+1)2\pi/3\rbrack \hat{e}_x 
	+ \sin \lbrack(n+1)2\pi/3\rbrack \hat{e}_y \nonumber\\
	\uveperp_n &\equiv -\sin \lbrack(n+1)2\pi/3\rbrack \hat{e}_x 
	+ \cos \lbrack(n+1)2\pi/3\rbrack \hat{e}_y \nonumber\\
\end{align}
with $n=1,2,3$ for the unit vectors along and perpendicular to the bonds, as shown in Fig.~\ref{FIG:Latt}b.

\begin{align}\label{EQ:kagomeD}
	\mathfrak{D}_{\vec{q}}^{\textrm{kagome}} =& k \sum_{n=1}^{6} \vec{B}_{n,\vec{q}}^{s}
	 \vec{B}_{n,-\vec{q}}^{s} \nonumber\\
	 &+ \frac{\kappa}{a^2} \sum_{n=1}^{3} \left( \vec{B}_{n,\vec{q}}^{b} 
	 +g\frac{ \vec{\Gamma}_{n,\vec{q}}^{b}+ \vec{\Gamma}_{n+3,\vec{q}}^{b}}{2} \right) \nonumber\\
	 &\quad * \left( \vec{B}_{n,-\vec{q}}^{b} 
	 +g\frac{ \vec{\Gamma}_{n,-\vec{q}}^{b}+ \vec{\Gamma}_{n+3,-\vec{q}}^{b}}{2} \right)	 ,
\end{align}
with
\begin{align}
	\vec{B}_{1,\vq}^{s} &= \{-\uvepara_{1},\uvepara_{1},0,0  \},
	\nonumber\\
	\vec{B}_{2,\vq}^{s} &= \{0,0,-\uvepara_{2},\uvepara_{2} \},
	\nonumber\\
	\vec{B}_{3,\vq}^{s} &= \{\uvepara_{3},0,0,-\uvepara_{3} \},
	\nonumber\\
	\vec{B}_{4,\vq}^{s} &= \{\uvepara_{1},
		-e^{2i \vq\cdot \uvepara_{1}}\uvepara_{1},0,0  \},
	\nonumber\\
	\vec{B}_{5,\vq}^{s} &= \{0,0,\uvepara_{2},
		-e^{2i \vq\cdot \uvepara_{2}}\uvepara_{2} \},
	\nonumber\\
	\vec{B}_{6,\vq}^{s} &= \{-e^{2i \vq\cdot \uvepara_{3}}\uvepara_{3},0,0,\uvepara_{3} \} ,
\end{align}
\begin{align}
	\vec{B}_{1,\vq}^{b} &= \{\uveperp_{1}, -\uveperp_{1}+\uveperp_{2},
	-e^{2i \vq\cdot \uvepara_{2}} \uveperp_{2}  \},
	\nonumber\\
	\vec{B}_{2,\vq}^{b} &= \{ -e^{2i \vq\cdot \uvepara_{3}} \uveperp_{3},
	\uveperp_{2}, -\uveperp_{2}+\uveperp_{3}\},
	\nonumber\\
	\vec{B}_{3,\vq}^{b} &= \{ -\uveperp_{3}+\uveperp_{1}, 
	 -e^{2i \vq\cdot \uvepara_{1}} \uveperp_{1}, \uveperp_{3}
	 \},
\end{align}
and
\begin{align}
	\vec{\Gamma}_{1,\vec{q}}^{b} &= \{\uveperp_{1}, -\uveperp_{1}-\uveperp_{2}, \uveperp_{2}  \}, \nonumber\\
	\vec{\Gamma}_{2,\vec{q}}^{b} &= \{\uveperp_{3}, \uveperp_{2}, -\uveperp_{2}-\uveperp_{3}  \}, \nonumber\\
	\vec{\Gamma}_{3,\vec{q}}^{b} &= \{ -\uveperp_{3}-\uveperp_{1}, \uveperp_{1}, \uveperp_{3}  \}, \nonumber\\
	\vec{\Gamma}_{4,\vec{q}}^{b} &= \{e^{-2i \vq\cdot \uvepara_{1}}\uveperp_{1}, -\uveperp_{1}-\uveperp_{2}, 
	e^{2i \vq\cdot \uvepara_{2}}\uveperp_{2}  \}, \nonumber\\
	% need to add those e^{} factors
	\vec{\Gamma}_{5,\vec{q}}^{b} &= \{e^{2i \vq\cdot \uvepara_{3}}\uveperp_{3}, 
	e^{-2i \vq\cdot \uvepara_{2}}\uveperp_{2}, -\uveperp_{2}-\uveperp_{3}  \}, \nonumber\\
	\vec{\Gamma}_{6,\vec{q}}^{b} &= \{ -\uveperp_{3}-\uveperp_{1}, e^{2i \vq\cdot \uvepara_{1}}\uveperp_{1}, 
	e^{-2i \vq\cdot \uvepara_{3}}\uveperp_{3}  \}, \nonumber\\
\end{align}
The unit vectors $\uvepara$ and $\uveperp$ are $2d$ vectors so the $\vec{B}$ vectors are $6$ dimensional.  The lattice constant of the kagome lattice is twice the bond length so it is $2a$.

The central force part of this dynamical matrix is the same as the nearest neighbor part of the dynamical matrix in Ref.~\cite{Mao2011a} as well as the central-force part in Ref.~\cite{Mao2013a}.

The Fourier transform on the kagome lattice follows similar rules defined for the hexagonal lattice, albeit with a different unit cell area $v_{0,k}=2\sqrt{3} a^2$.

%\bibliography{isostaticity2}

\end{document}